\documentclass[useAMS,usenatbib]{mn2e}

\usepackage{graphicx}
\usepackage{subfigure}
\usepackage{amssymb}
\usepackage{amsmath}
\usepackage{tabularx}
\usepackage{float}
\usepackage{placeins}
\usepackage{adjustbox}

\title[GMC Mass Dependence of Radiative Feedback]{Simulating radiative feedback and star cluster formation in GMCs: II. Mass dependence of cloud destruction and cluster properties}

\author[C.S.\ Howard, R.E.\ Pudritz, \& W.E.\ Harris ]{Corey \ S. \ Howard$^{1}$\thanks{E-mail: howardcs@mcmaster.ca}, Ralph \ E.\ Pudritz$^{1,2}$, William \ E.\ Harris$^{1}$\\
$^{1}$Department of Physics and Astronomy, McMaster University, 1280 Main St.~W, Hamilton, ON L8S 4M1, Canada\\
$^{2}$Origins Institute, McMaster University, 1280 Main St.~W, Hamilton, ON L8S 4M1, Canada
}

\begin{document}
\bibliographystyle{mn2e}

\date{13 March 2017}

\pagerange{\pageref{firstpage}--\pageref{lastpage}} \pubyear{2017}

\maketitle

\label{firstpage}

\begin{abstract}

\noindent The process of radiative feedback in Giant Molecular Clouds (GMCs) is an important mechanism for limiting star cluster formation through the heating and 
ionization of the surrounding gas. We explore the degree to which radiative feedback affects early ($\lesssim$5 Myr) cluster formation in GMCs having masses that range 
from 10$^{4-6}$ M$_{\odot}$ using the \textit{FLASH} code. The inclusion of radiative feedback lowers the efficiency of cluster formation by 20-50\% relative to hydrodynamic 
simulations. Two models in particular --- 5$\times$10$^4$ and 10$^5$ M$_{\odot}$ --- show the largest suppression of the cluster formation efficiency, corresponding to a 
factor of $\sim$2. For these clouds only, the internal energy, a measure of the energy injected by radiative feedback, exceeds the gravitational potential for a significant 
amount of time. We find a clear relation between the maximum cluster mass, M$_{cl,max}$, formed in a GMC of mass M$_{GMC}$; M$_{cl,max}\propto$ M$_{GMC}^{0.81}$. This scaling 
result suggests that young globular clusters at the necessary scale of $10^6 M_{\odot}$ form within host GMCs of masses near $\sim 5 \times 10^7 M_{\odot}$. We compare 
simulated cluster mass distributions to the observed embedded cluster mass function ($dlog(N)/dlog(M) \propto M^{\beta}$ where $\beta$ = -1) and find good agreement 
($\beta$ = -0.99$\pm$0.14) only for simulations including radiative feedback, indicating this process is important in controlling the growth of young clusters. However, 
the high star formation efficiencies, which range from 16-21\%, and high star formation rates compared to locally observed regions suggest other feedback mechanisms are 
also important during the formation and growth of stellar clusters.

\end{abstract}

\begin{keywords}
galaxies: star clusters: general -- H ii regions -- radiative transfer -- stars: formation -- methods: numerical -- hydrodynamics 
\end{keywords}

\section{Introduction}

\indent \indent The formation of star clusters takes place within dense (n $>$ 10$^4$ cm$^{-3}$) clumps of molecular gas embedded in Giant Molecular Clouds (GMCs) \citep{Lada2003,MaclowKlessen,BertoldiMckee,Kruij2012}.
These clouds are supersonically turbulent and highly filamentary with the most massive clusters forming at the intersection of these filaments \citep{Balsara,Banerjee2006,Schneider2012,Kirk2013}. Since star clusters are almost 
exclusively formed in GMCs, understanding the processes that lead to their formation and destruction is vital for a complete understanding of galaxy evolution over cosmic time.

The properties of GMC within a galaxy --- such as the mass and virial parameter \citep{Solomon1987,Rosolowsky2007,Hernandez2015, Howard1} --- vary from cloud to cloud. Within
the Milky Way (MW), the typical size of a GMC ranges from 50 pc to several hundreds of parsecs with masses in the range of $\sim$10$^{4-7}$ M$_{\odot}$ \citep{Fukui}. More specifically,
the mass distribution of clouds within the inner disk of the MW follows a power law $dN/dM \propto M^{\alpha}$ where $\alpha$ $\sim$ -1.5 \citep{Sanders,Solomon1987,Roso2005}. The power law 
index for the GMC mass distribution in other Local Group galaxies is found to be significantly steeper, ranging from -1.7 for the LMC to -2.5 for M33 \citep{Blitz,Roso2005}.

The mass of a GMC has a direct impact on a cluster that form within it. Both simulations \citep{Fujii} and observations \citep{Hughes} indicate a relation between the mass of a GMC (M$_{GMC}$)
and the maximum mass cluster (M$_{c,max}$) it produces of the form M$_{c,max}\propto$M$_{GMC}^{0.5}$. Based on the similarity of the mass scaling of GMCs and star clusters, \citet{HP1994}
proposed that Globular Clusters (GCs) originated in Supergiant Molecular Clouds ($\geq$10$^{7}$ M$_{\odot}$). Overall, these results suggest that the massive stellar content should increase with GMC mass. This is indeed borne out in
observations of the LMC \citep{Kawamura, FukuiARAA} which show that GMCs with large HII regions, indicating the presence of massive stars, are typically more massive than GMCs 
with no, or low luminosity, HII regions. 


The overall conversion of molecular gas into stars, regardless of cloud mass, is an inherently inefficient process. Typical estimates of the star formation efficiency over the lifetime of an individual
GMC in the MW range from 2-5\% \citep{Lada2003, McKee2007,Murray2011}. 

The question of what limits star formation in a GMC to such low values, and ultimately disrupts the cloud, is debated. Both turbulence \citep{Klessen2000,Bate2003,Bonnell2008} and magnetic fields \citep{MyersGoodman1988,Tilley2007,Federrath2012} can provide added pressure
support against gravitational collapse and lower the star formation rate per freefall time, but cannot completely disperse the GMC. Alternatively, feedback from newly-formed
stars can both limit the star formation efficiency and destroy the GMC via the input of energy and momentum into the gas. 

The goal of this paper is to explore how cluster formation and radiative feedback affect GMCs and ultimately star cluster properties. For this purpose, we present the results from a suite of simulations which examine the role of radiative feedback in 5 clouds ranging from 10$^{4-6}$ M$_{\odot}$. The initial 
average density and the initial virial parameter are identical for all models in order to ensure all observed differences are due solely to varying the mass.

Stellar feedback comes in many forms --- protostellar jets \citep{Li2006,Maury2009,Federrath2014}, stellar winds \citep{Dale2008,Gatto2017}, ionization/heating of the gas \citep{Dale2005,Peters2010,Klassen2012}, radiation pressure \citep{Krumholz2012,Murray2010},
and supernovae feedback \citep{Rogers,Fierlinger,Keller2014,Gatto2015,Walch2015}. Of these mechanism, radiative feedback has been suggested as being most important during the early phases of cluster formation, 
particularly in clusters which are hosting massive star formation \citep{Whitworth79,Matzner2002,Murray2010,Dale2012,Bate2012}. The heating and ionization of the gas surrounding star-forming clusters prevents further 
fragmentation, and expanding HII regions can drive further turbulence \citep{Grit2009,Boneberg}. Direct radiation pressure from high energy photons interacting with dust grains can also
drive strong outflows.

Previous studies which examine the impact of radiative feedback on both small (individual cluster) scales and large (entire GMC) scales show that the overall star formation 
efficiency can be reduced \citep{Dale2007,Peters2010,Dale2012,Bate2012,Klassen2012,Walch2012}. In particular, the work of \citet{Dale2012} showed that radiative feedback 
produces large scale HII regions which drive significant gas outflows from the cloud. This is especially important in low mass ($\sim$10$^4$ M$_{\odot}$) clouds. Despite the production 
of these large features, the influence on star formation efficiencies and rates was small. Their models, however, were limited to gravitationally bound clouds.

Our own work \citep{Howard1} also showed that the inclusion of radiative feedback
did reduce the efficiency of cluster formation, but only by a maximum of $\sim$10\%. This study was limited to a single GMC mass (10$^6$ M$_{\odot}$) which, while present in 
the MW, are not typical of the average GMC as illustrated by the powerlaw mass distribution discussed above. Moreover, since the properties of the population of clusters
formed in a GMC depends on its initial mass, the effects of radiative feedback can possibly differ when considering a spectrum of cloud masses.

We evolve all models to $\sim$5 Myr, at which point supernovae are expected to become a significant factor in the cloud's evolution. To make this computationally feasible, we
make use of sink particles to represent star clusters in combination with a custom subgrid model to follow the formation of stars within the cluster. We discuss the details of this
model, our numerical methods, and the GMC initial conditions in Section 2. 

\begin{table*}
\centering
\begin{adjustbox}{width=1\textwidth}
\small
\begin{tabular}{|c|c|c|c|c|c|c|}
\hline
\textbf{Mass (M$_{\odot}$)} & \textbf{Radius (pc)} & \textbf{Virial Parameter} & \textbf{Initial Mach Number} & \textbf{Resolution (pc)} & \textbf{Particle Radius (pc)} & \textbf{Radiative Threshold (M$_{\odot}$)} \\ \hline
10$^4$ & 7.67 & 3 & 13.6 & 0.13 & 0.33 & 100  \\ \hline
5$\times$10$^4$ & 13.1 & 3 & 23.3 & 0.23 & 0.58 & 100  \\ \hline
10$^5$ & 16.5 & 3 & 36.2 & 0.29 & 0.73 & 1000  \\ \hline
5$\times$10$^5$ & 28.3 & 3 & 50.3 & 0.25 & 0.62 & 1000  \\ \hline
10$^6$ & 33.8 & 3 & 73.1 & 0.31 & 0.78 & 1000 \\ \hline
\end{tabular}
\end{adjustbox}
\caption{Summary of parameters for each simulation. Note that two simulations were completed for every entry in the table --- one including radiative feedback and one without radiative feedback.}
\end{table*}

In Section 3, we discuss the global evolution of our cloud models and the role that radiative feedback
plays on the final cluster and star formation efficiencies. We find that feedback reduces these efficiencies for all clouds, but it is most significant in the 5$\times$10$^4$ and 10$^5$ M$_{\odot}$ clouds 
which have the efficiency of cluster formation reduced by approximately a factor of 2. We show that this is the result of a trade off between the energy injected by radiative feedback  and the
gravitational potential energy of the cloud. GMCs in this particular mass range are massive enough to form a population of massive stars but have a small enough gravitational potential to become
unbound under the influence of radiative feedback.

In Section 4, we compare our star formation rates and initial cluster mass function to their observed counterparts. We find that the slope of our cluster mass
function over the range of masses observed for embedded clusters is consistent with observations only when radiative feedback is included. However, the combination of
high SFRs at late times and star formation efficiencies which range between 16 and 21\% suggest that radiative feedback alone is not responsible for limiting early 
star and cluster formation.   

\section{Numerical Methods} \label{methods}

\indent \indent Here, we provide a brief description of the numerical methods employed in our simulations. For more detail, we refer the reader to \citet{Howard1}.

We perform numerical simulations using the Adaptive Mesh Refinement (AMR) code FLASH (version 2.5) \citep{Fryxell2000} which 
solves the compressible gas dynamic equations on a Eulerian grid. FLASH also includes modules to treat self-gravity, radiative transfer, star formation, and cooling via molecules and dust. 

Gas cooling is treated using the methods from \citet{Banerjee+2006} which handles cooling via gas-dust interactions, H$_{2}$ dissociation, and molecular line emission. The cooling
rates for molecular line emission and gas-dust transfer are taken from \citet{Neufeld} and \citet{Goldsmith}, respectively.

\begin{figure*}
\begin{tabular}{ccc}
\includegraphics[width=0.33\linewidth]{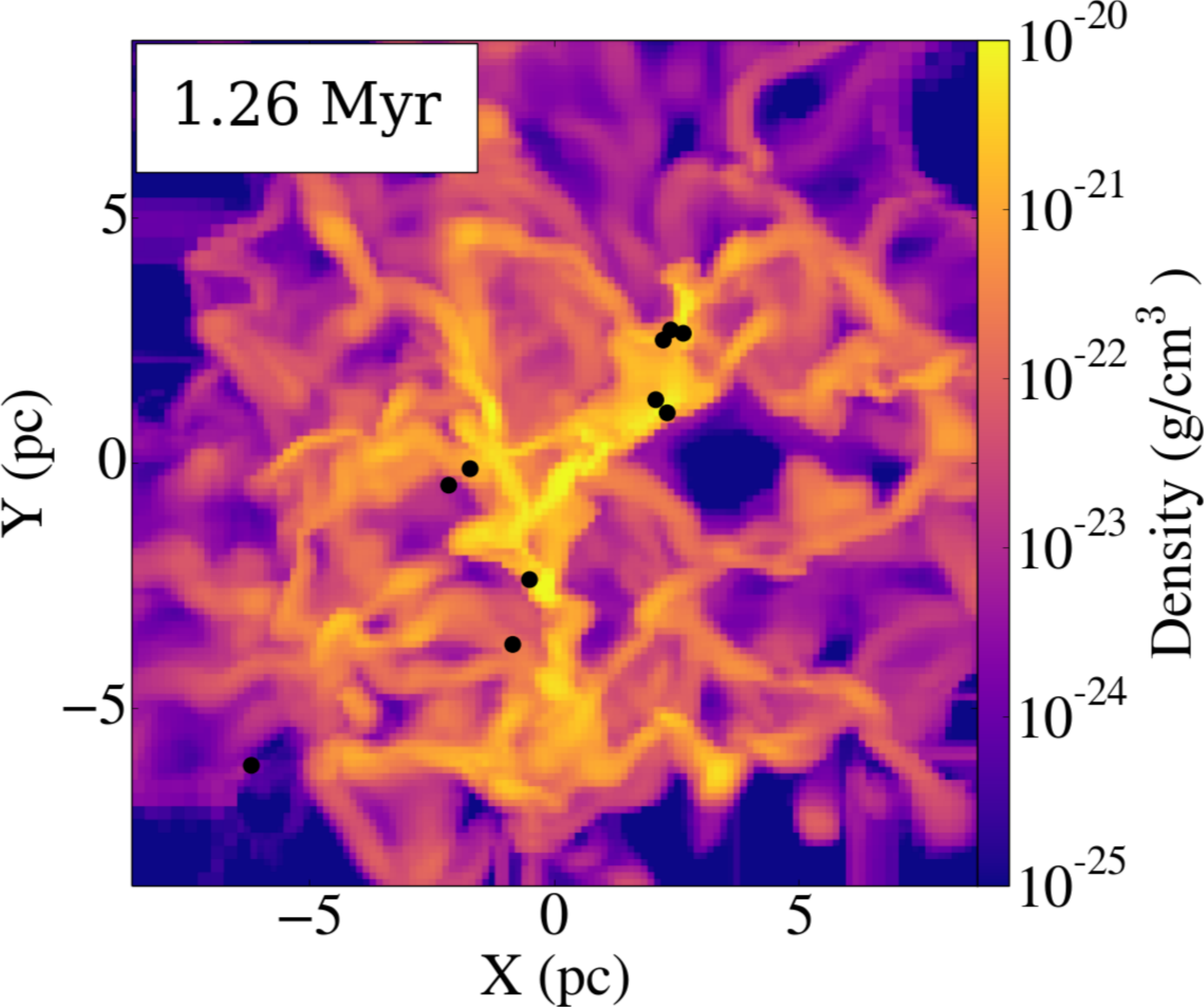} & \includegraphics[width=0.33\linewidth]{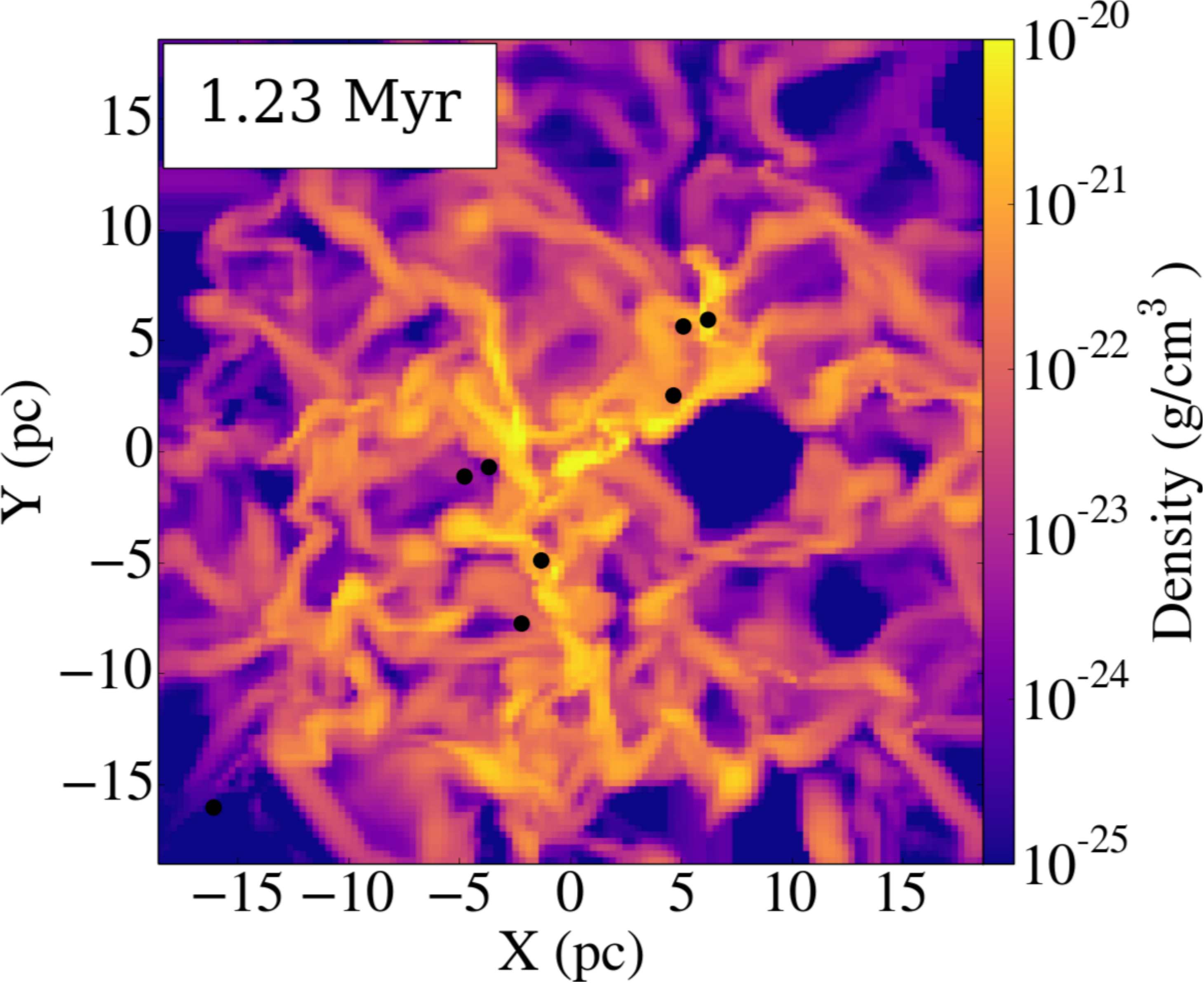} & \includegraphics[width=0.33\linewidth]{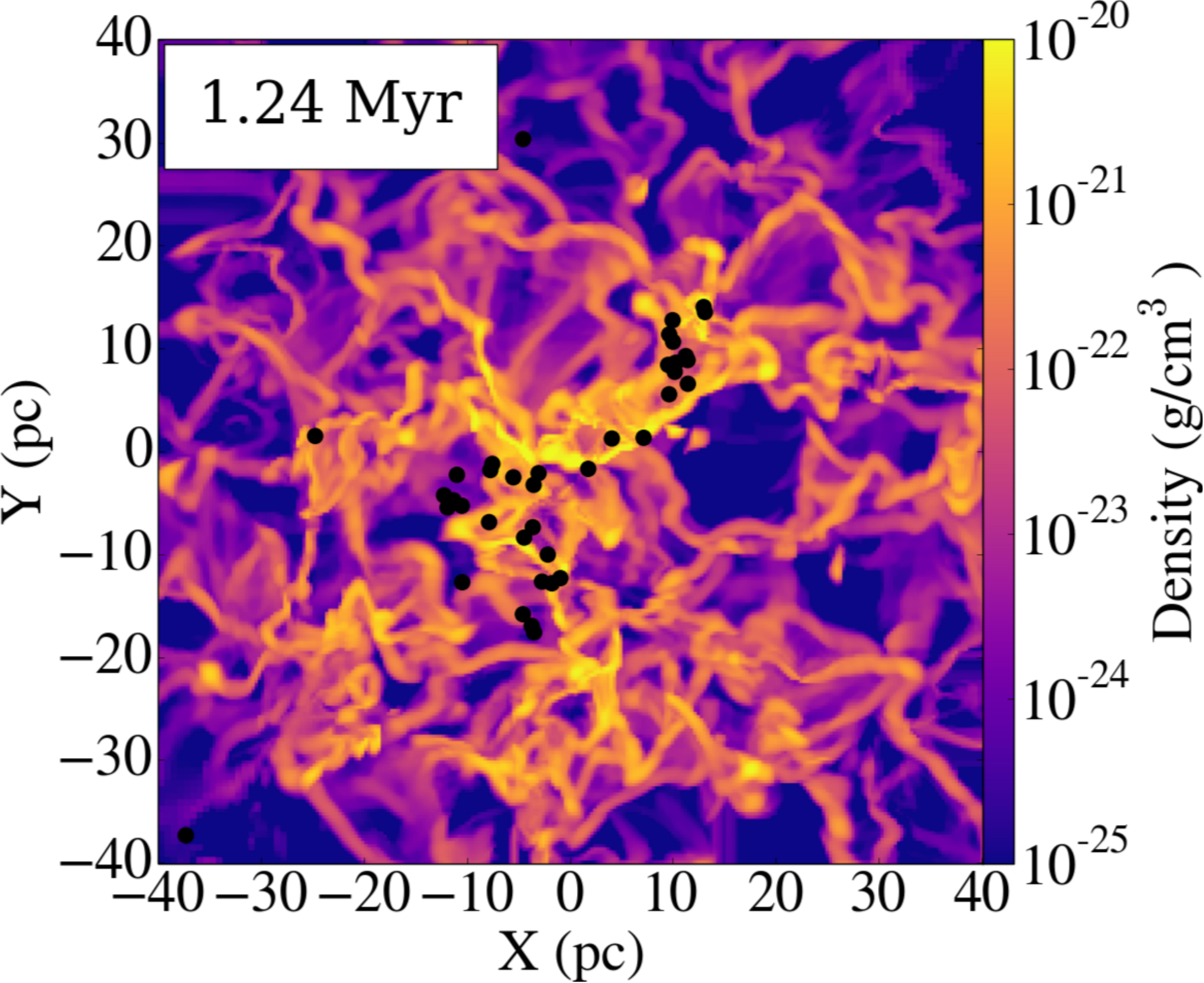} \\
\includegraphics[width=0.33\linewidth]{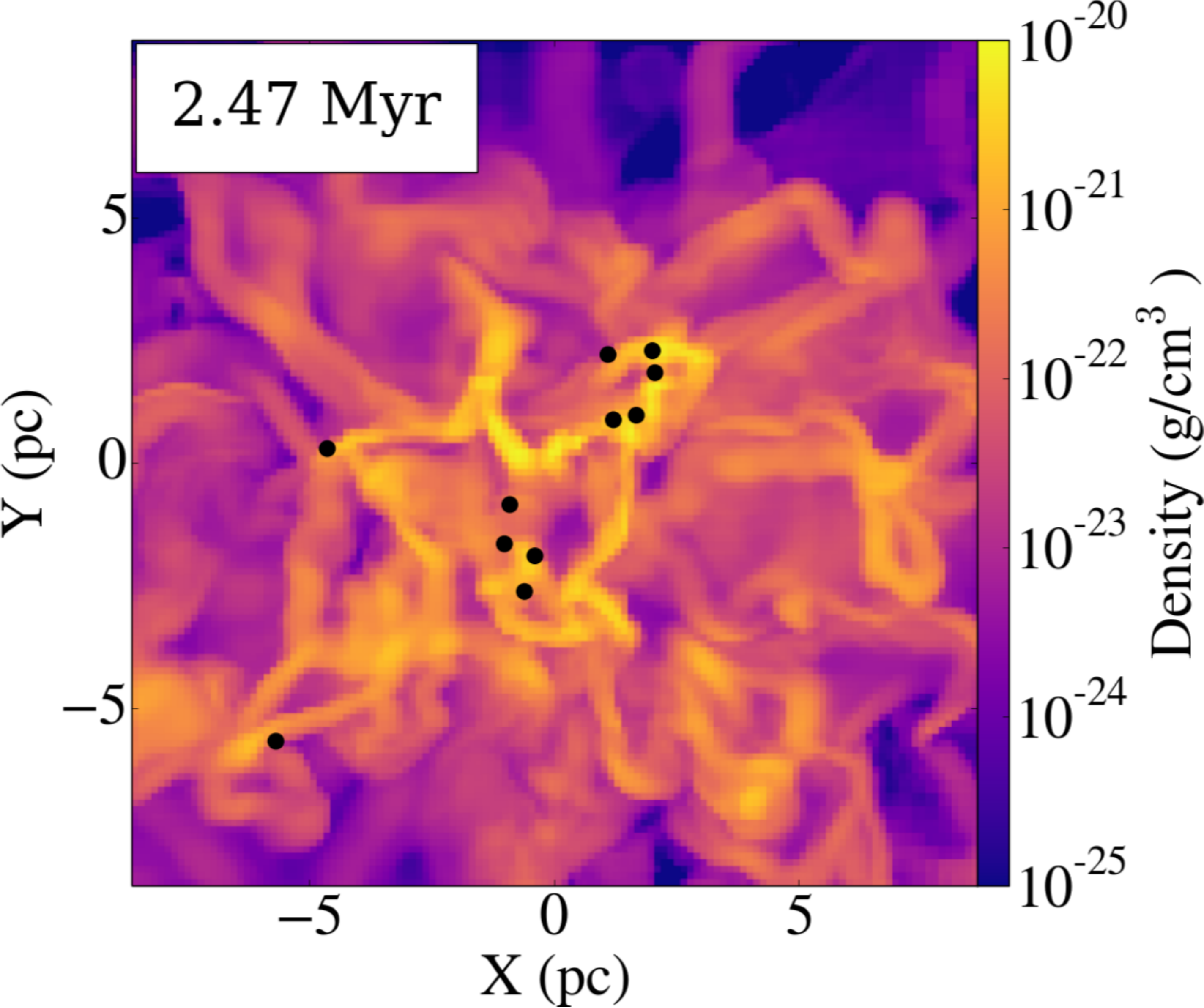} & \includegraphics[width=0.33\linewidth]{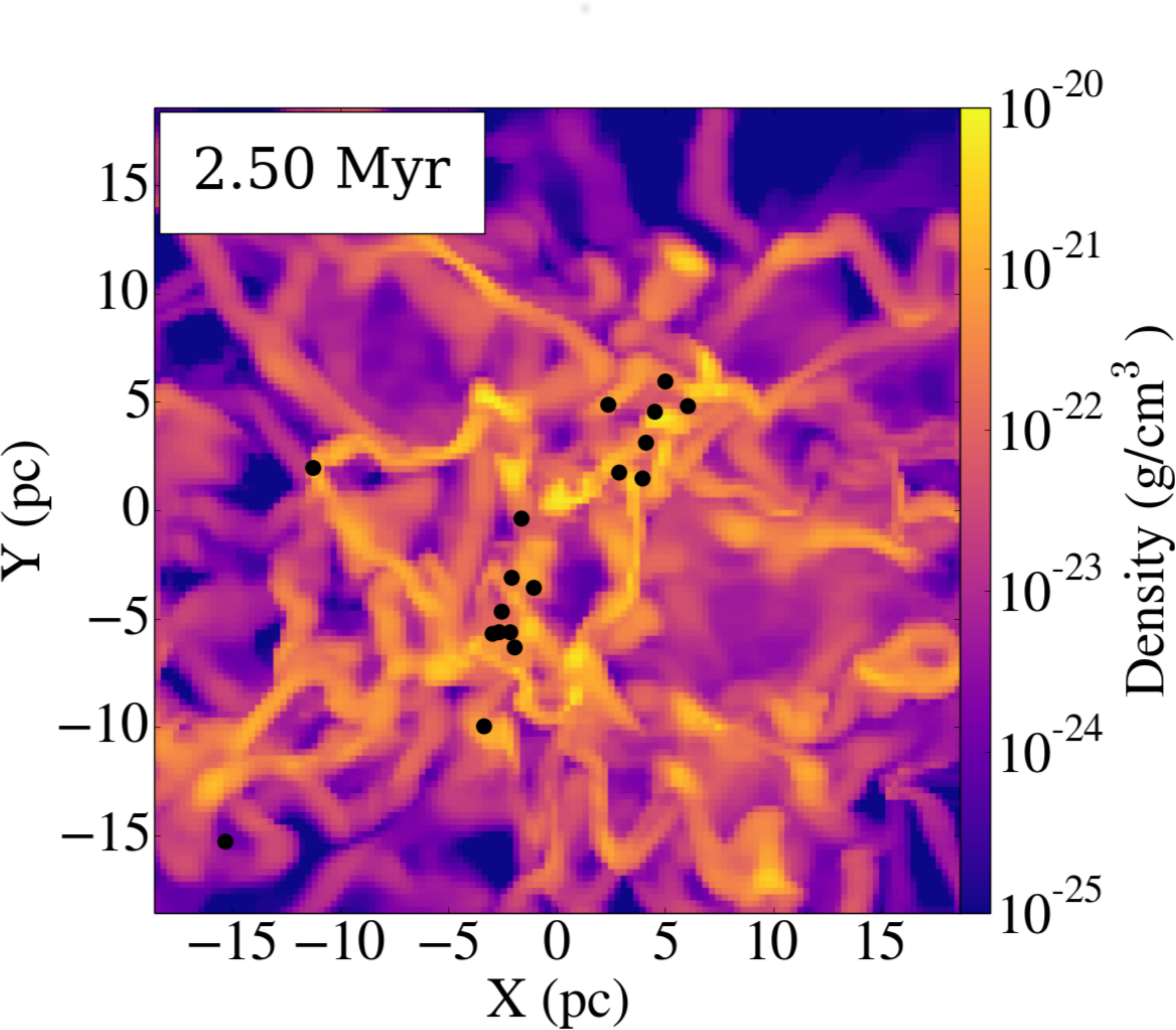} & \includegraphics[width=0.33\linewidth]{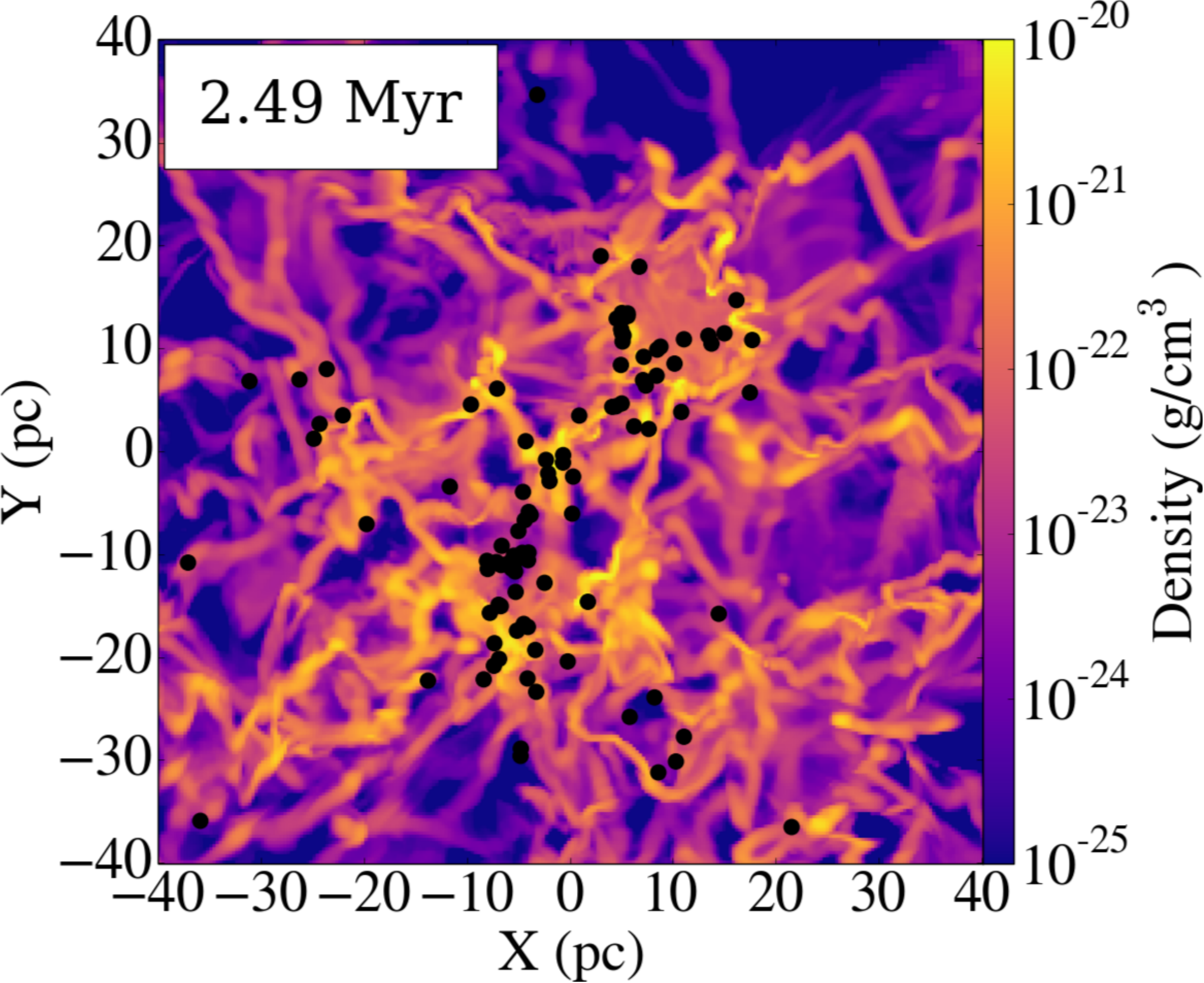} \\
\includegraphics[width=0.33\linewidth]{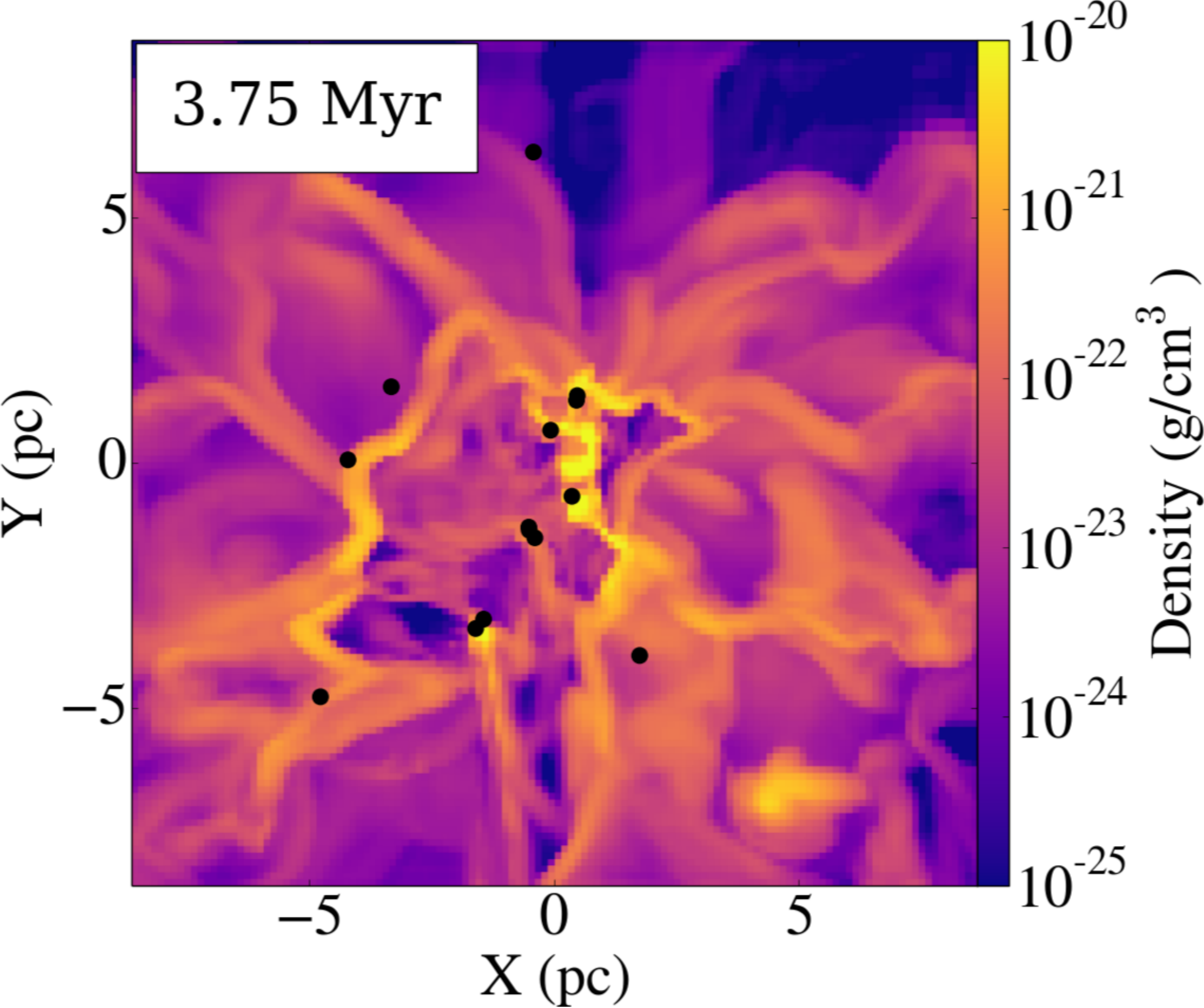} & \includegraphics[width=0.33\linewidth]{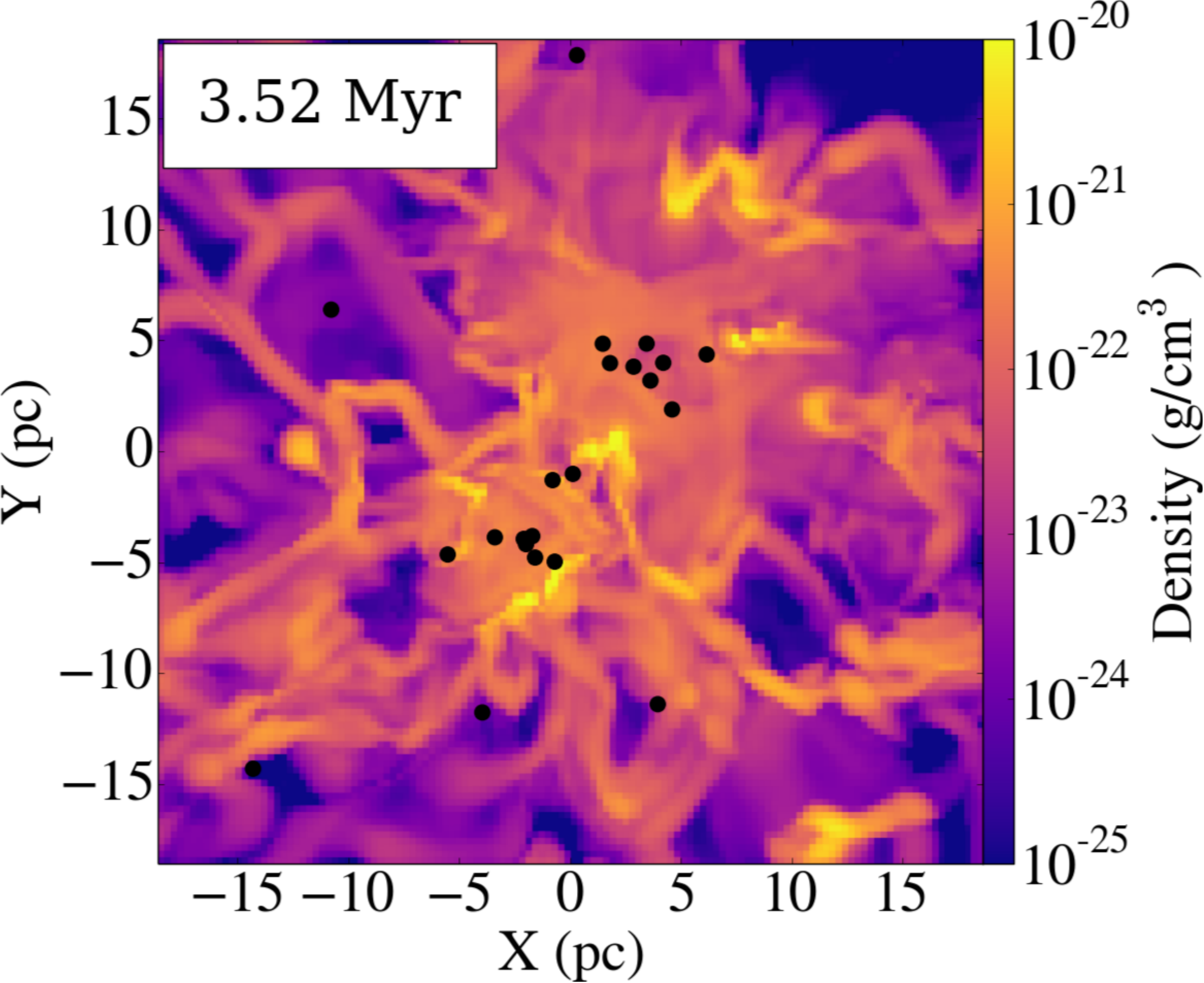} & \includegraphics[width=0.33\linewidth]{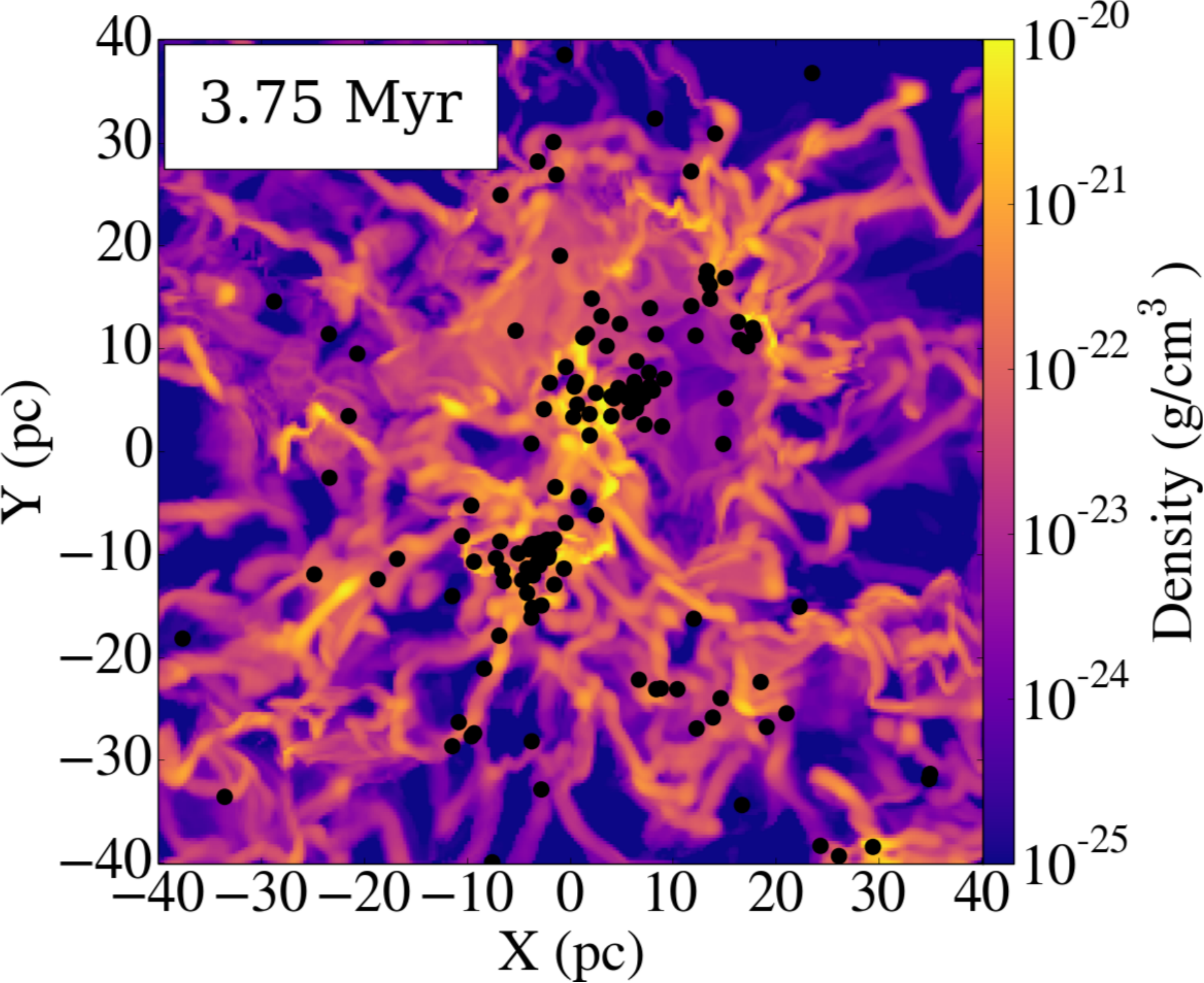} \\
\includegraphics[width=0.33\linewidth]{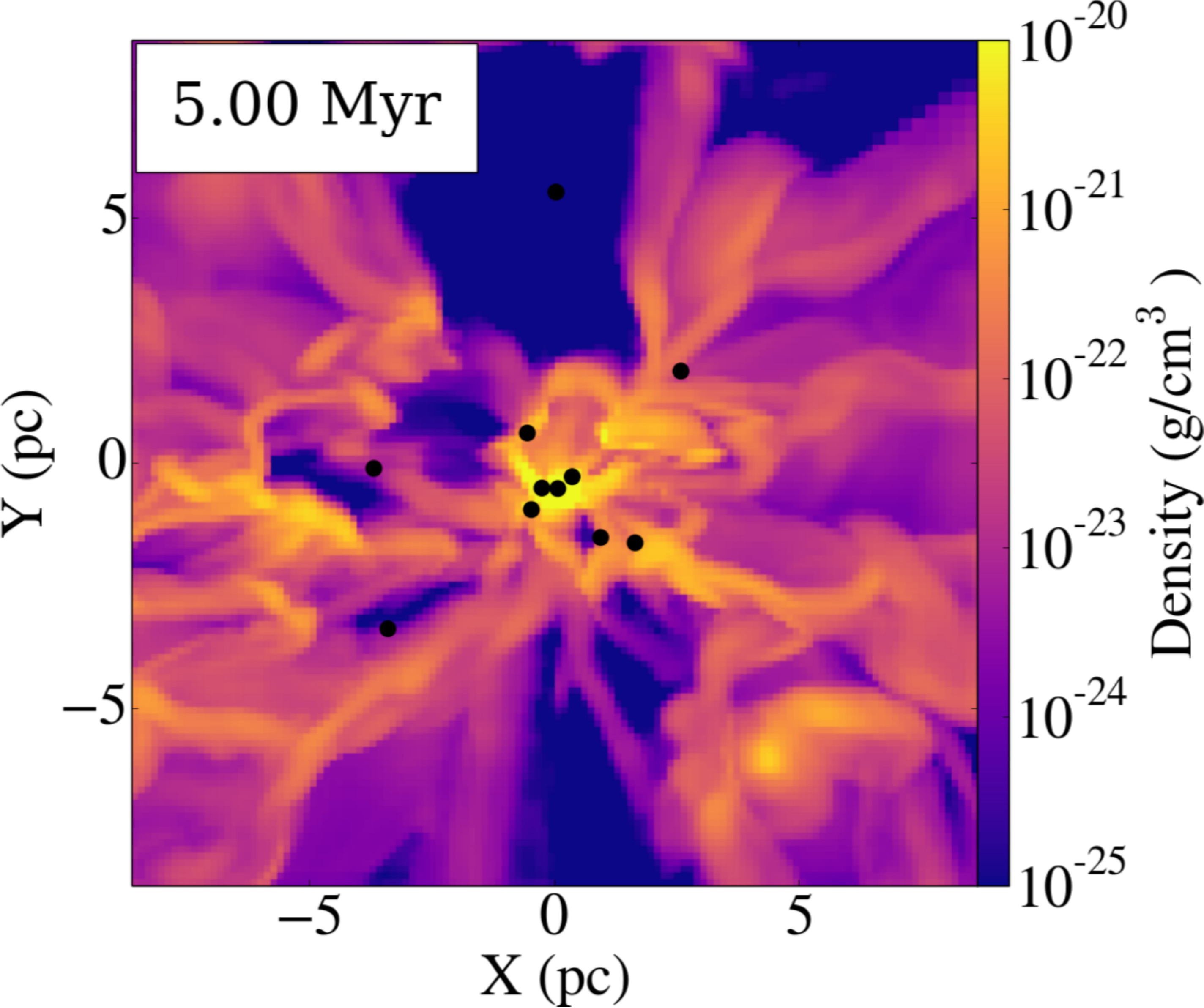} & \includegraphics[width=0.33\linewidth]{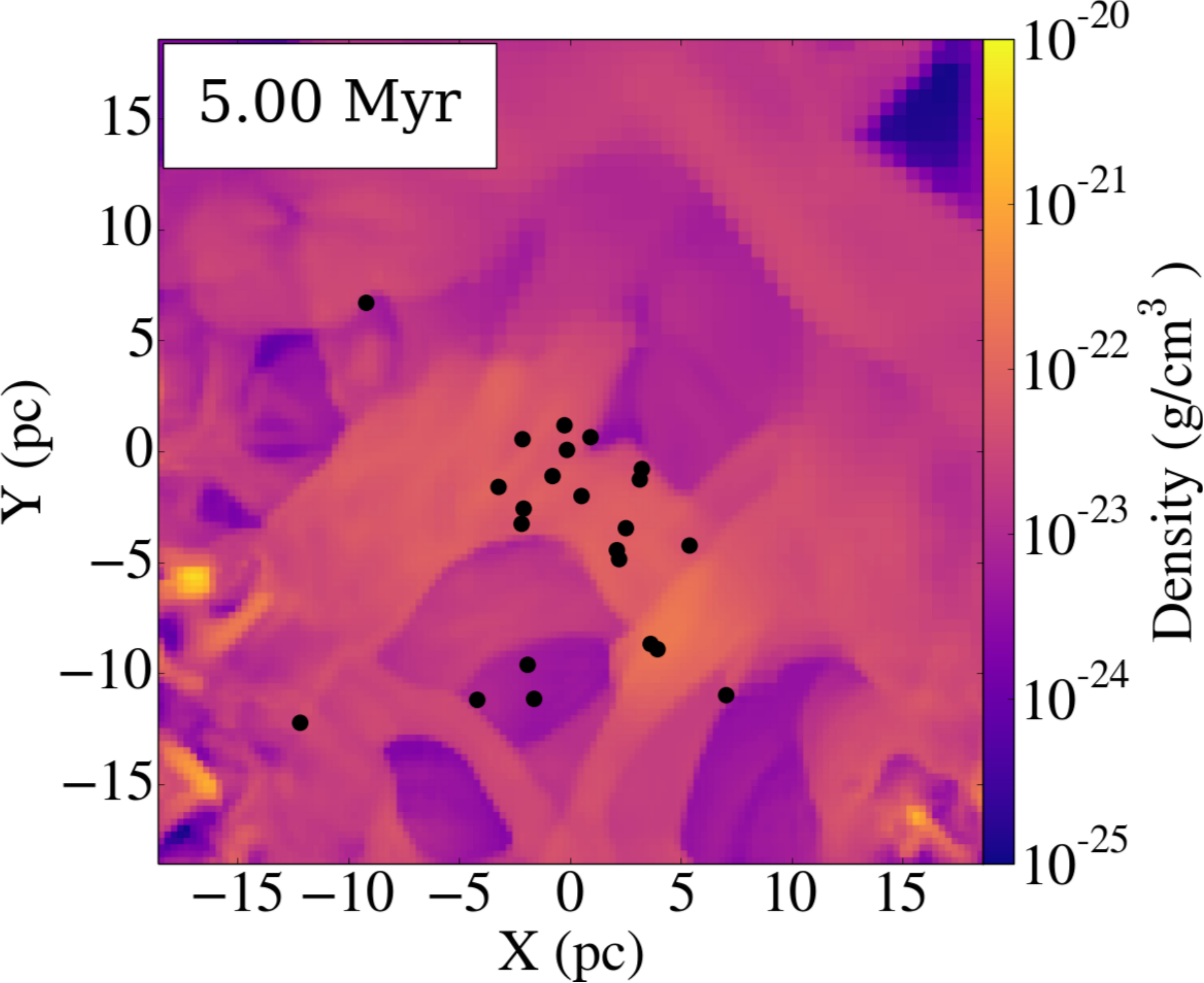} & \includegraphics[width=0.33\linewidth]{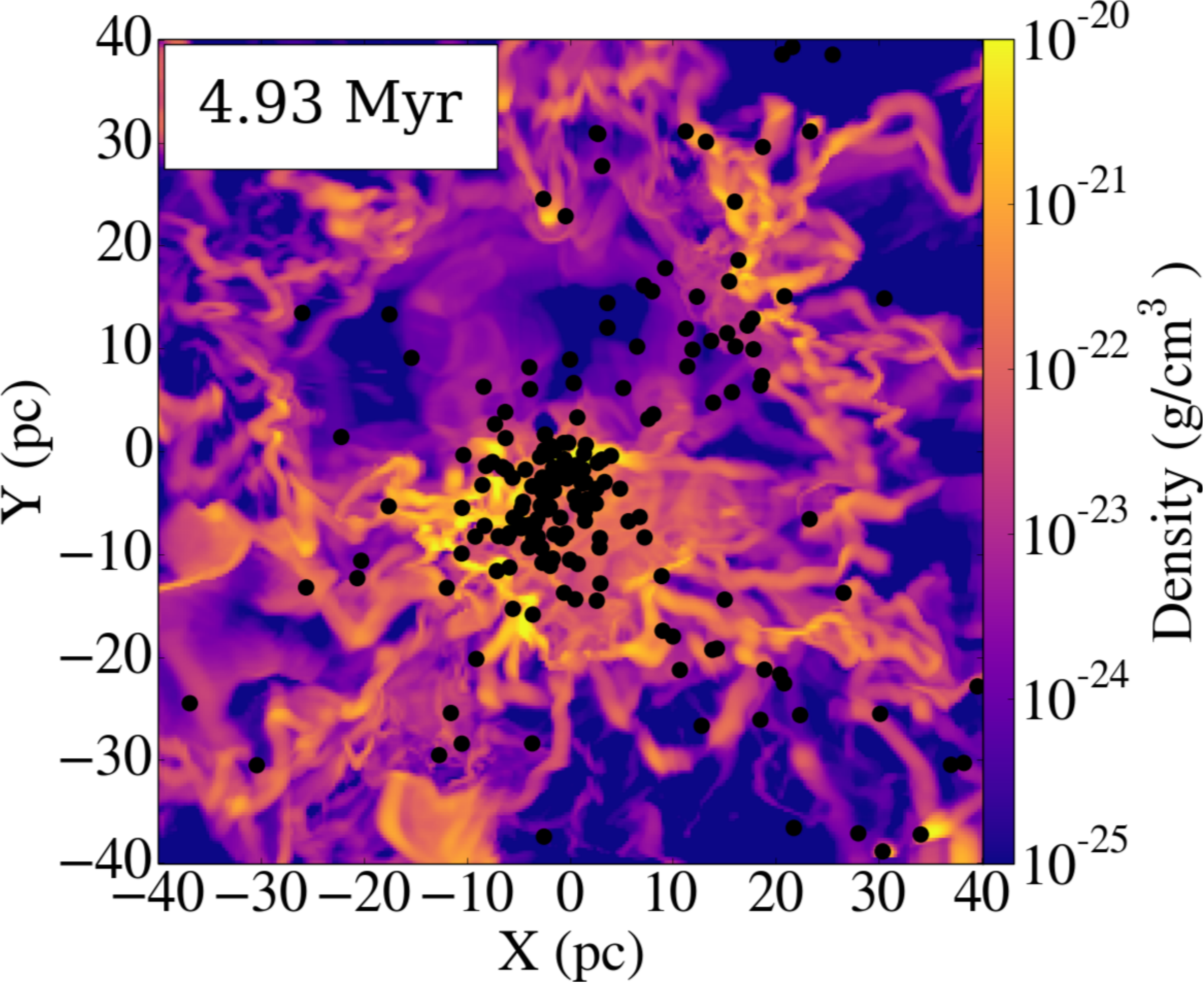} \\
\end{tabular}
\caption{Density slices through the center of the simulation volume for the 10$^4$ (left), 10$^5$ (center), and 10$^6$ (right) M$_{\odot}$ GMCs. Time, shown in the top left
of each panel, increases from top to bottom. Cluster locations are projected onto this slice and shown by black circles. Note that the physical (xy) scales change with 
cloud mass (10x10 pc, 20x20 pc, and 40x40 pc from left to right).}
\label{fig:slice_dens}
\end{figure*}

\begin{figure*}
\begin{tabular}{ccc}
\includegraphics[width=0.33\linewidth]{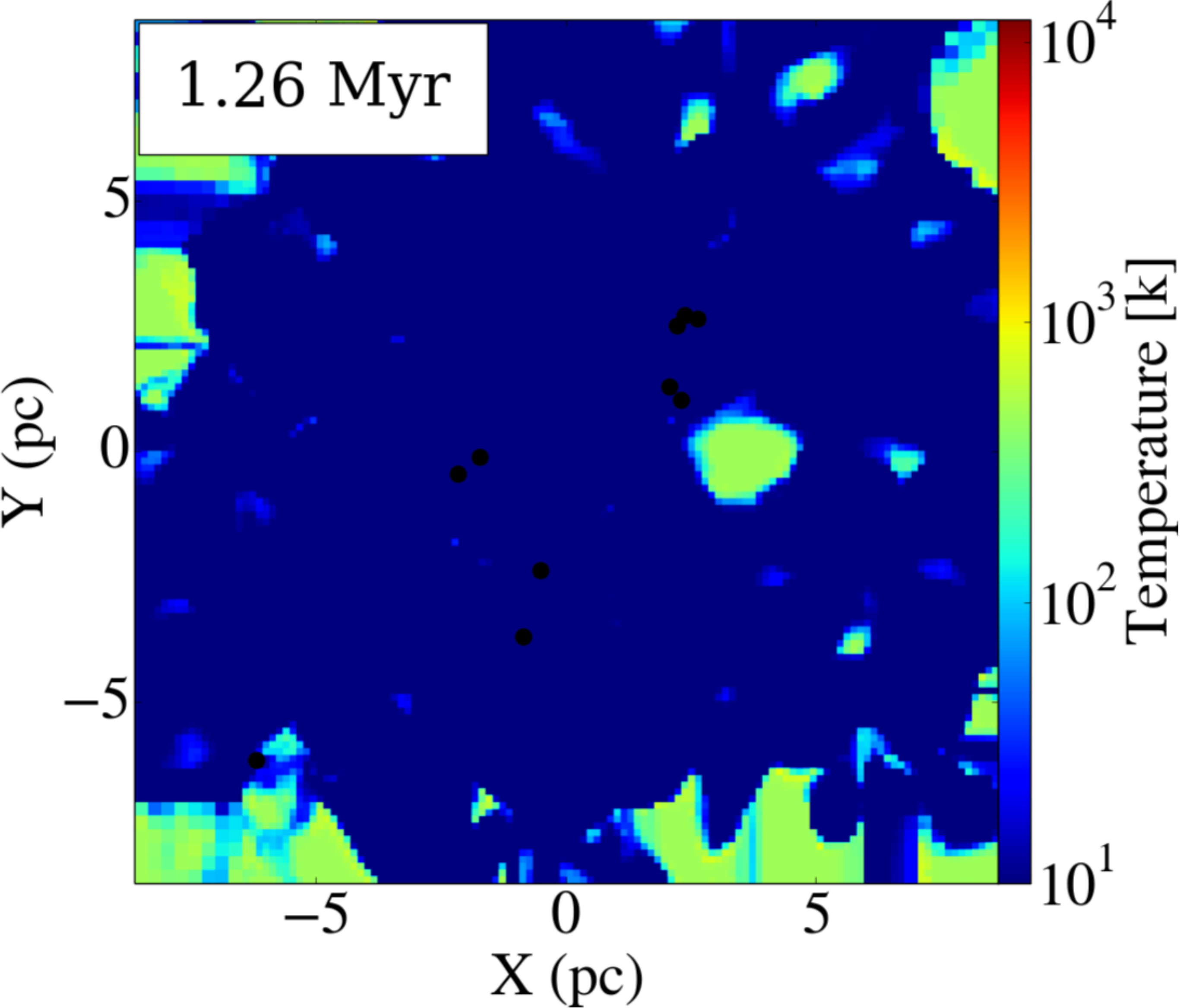} & \includegraphics[width=0.33\linewidth]{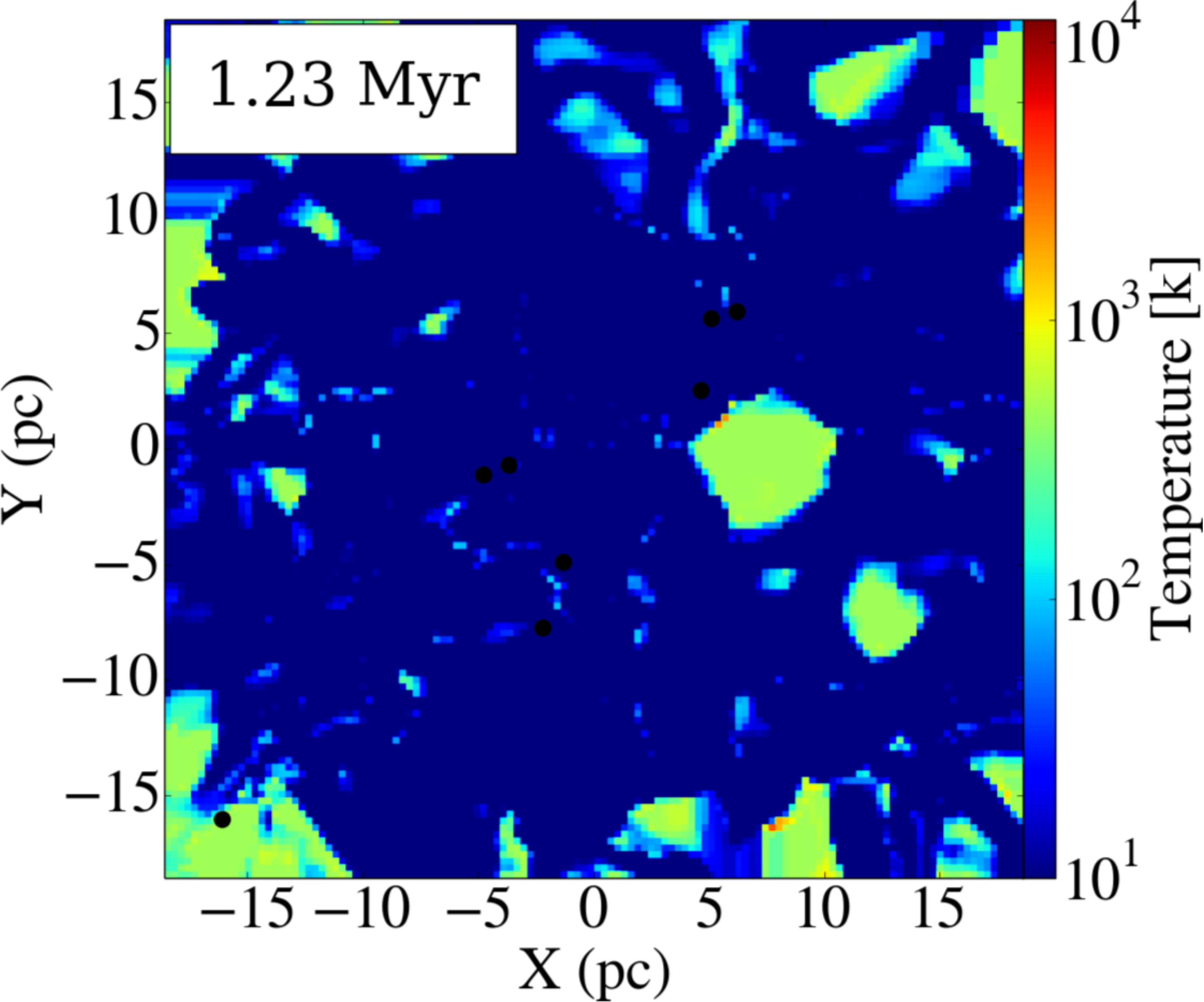} & \includegraphics[width=0.33\linewidth]{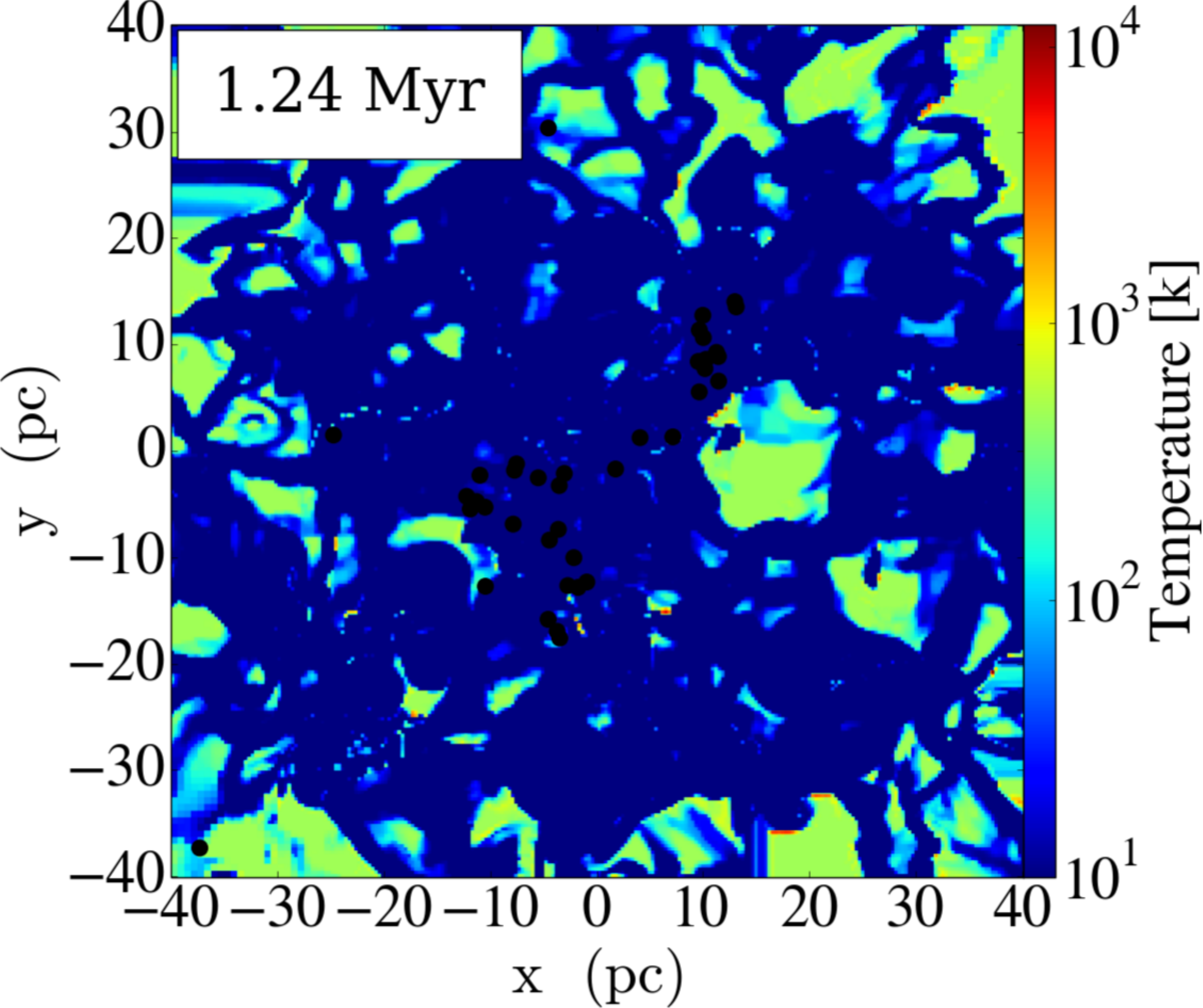} \\
\includegraphics[width=0.33\linewidth]{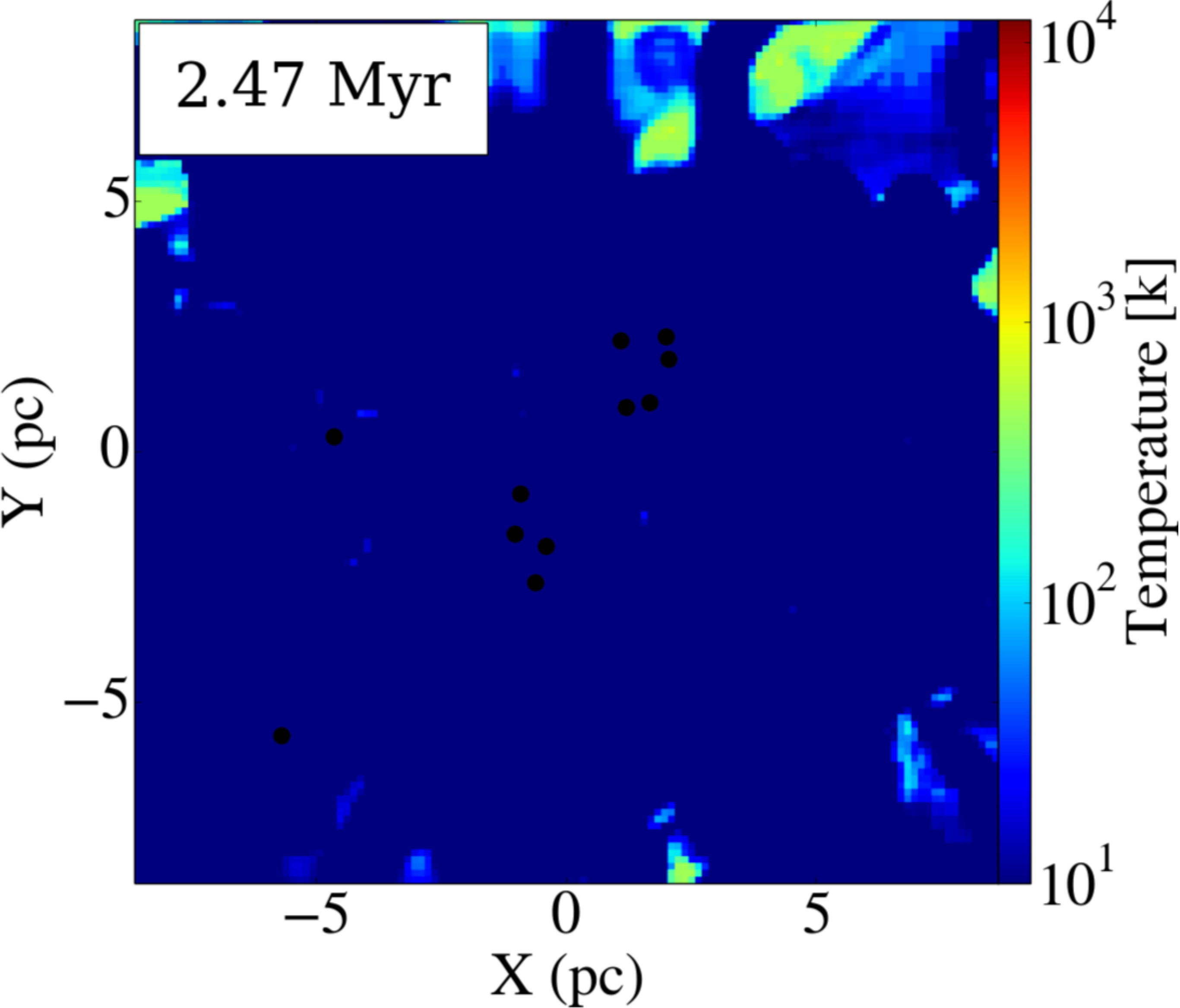} & \includegraphics[width=0.33\linewidth]{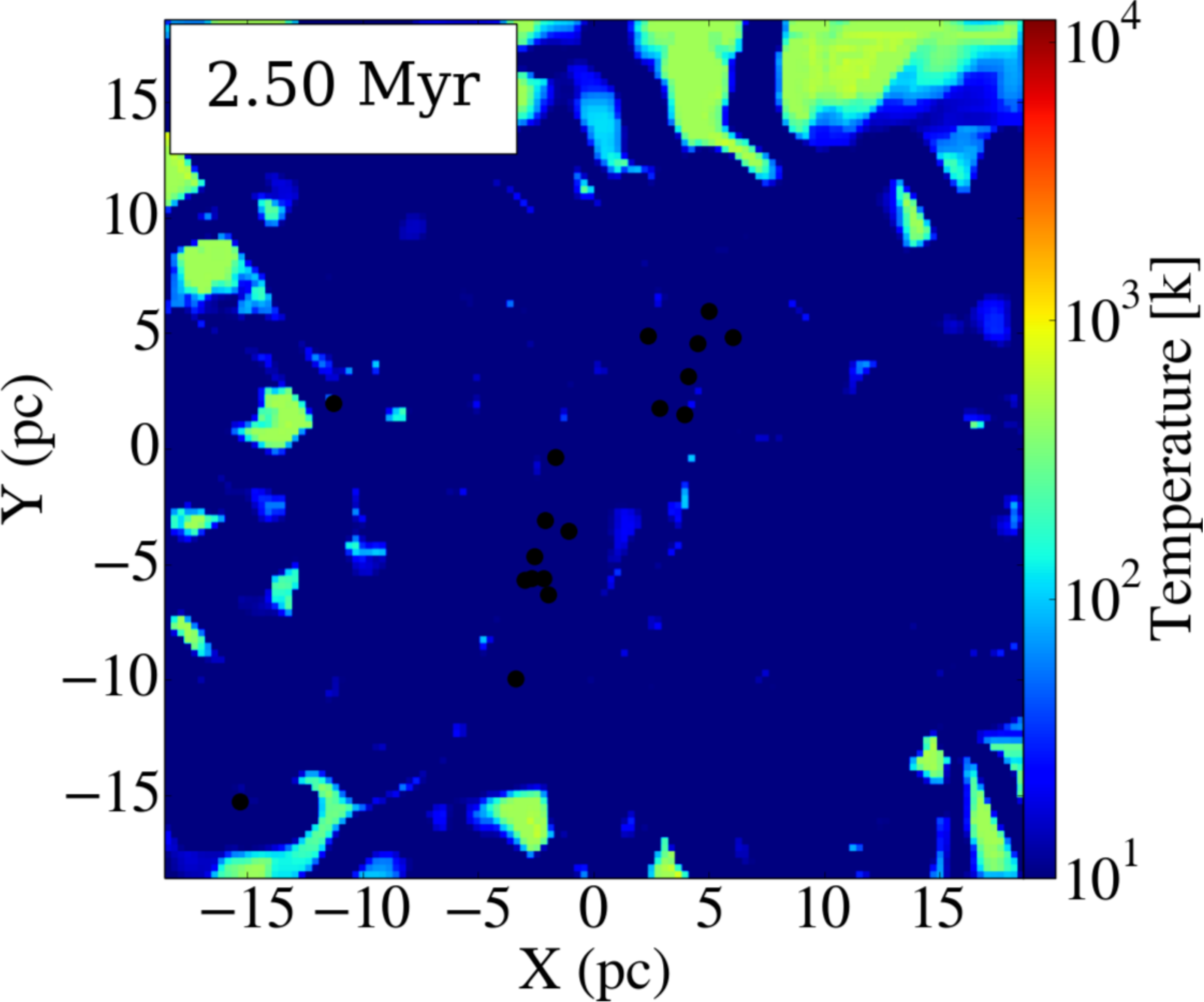} & \includegraphics[width=0.33\linewidth]{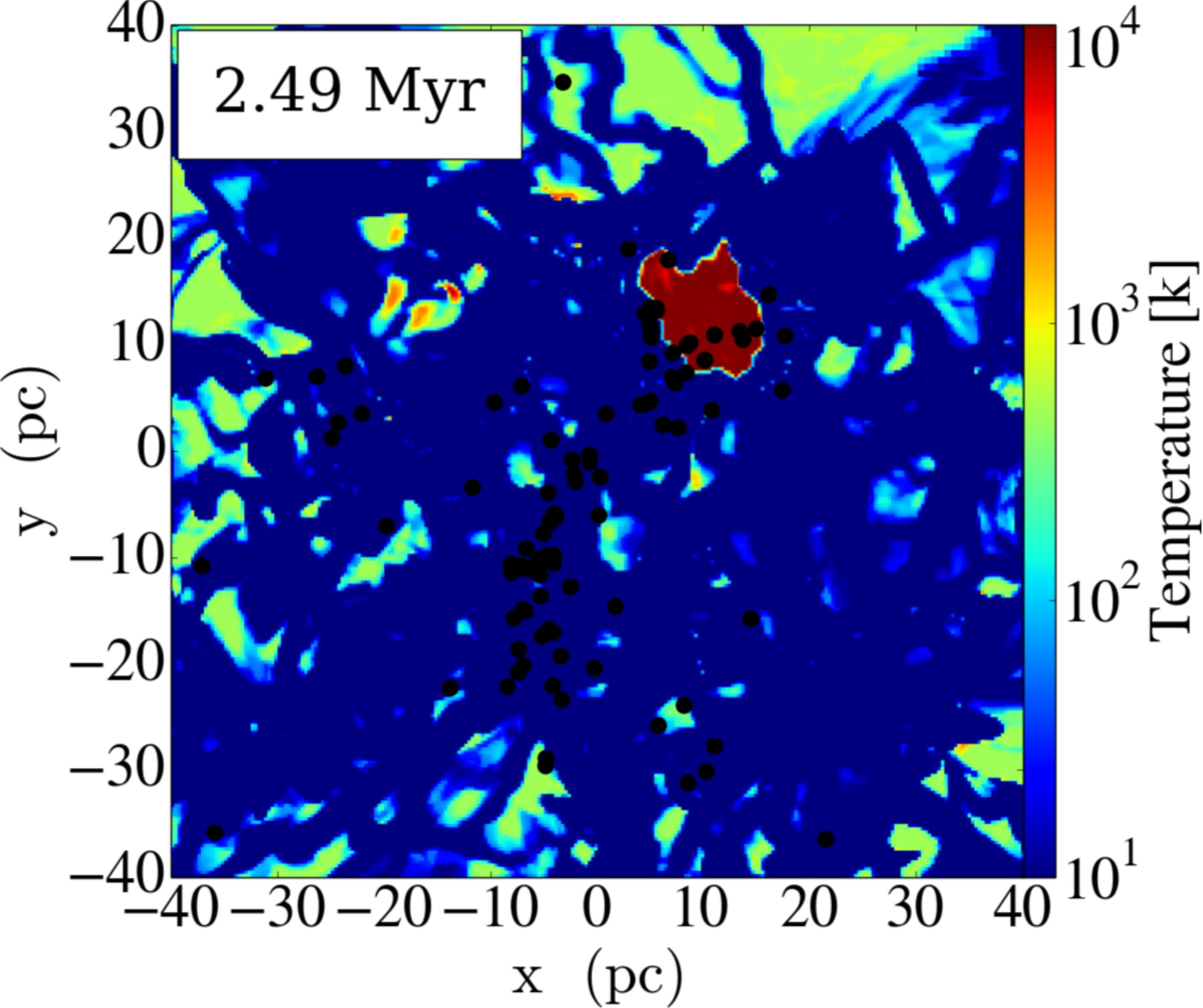} \\
\includegraphics[width=0.33\linewidth]{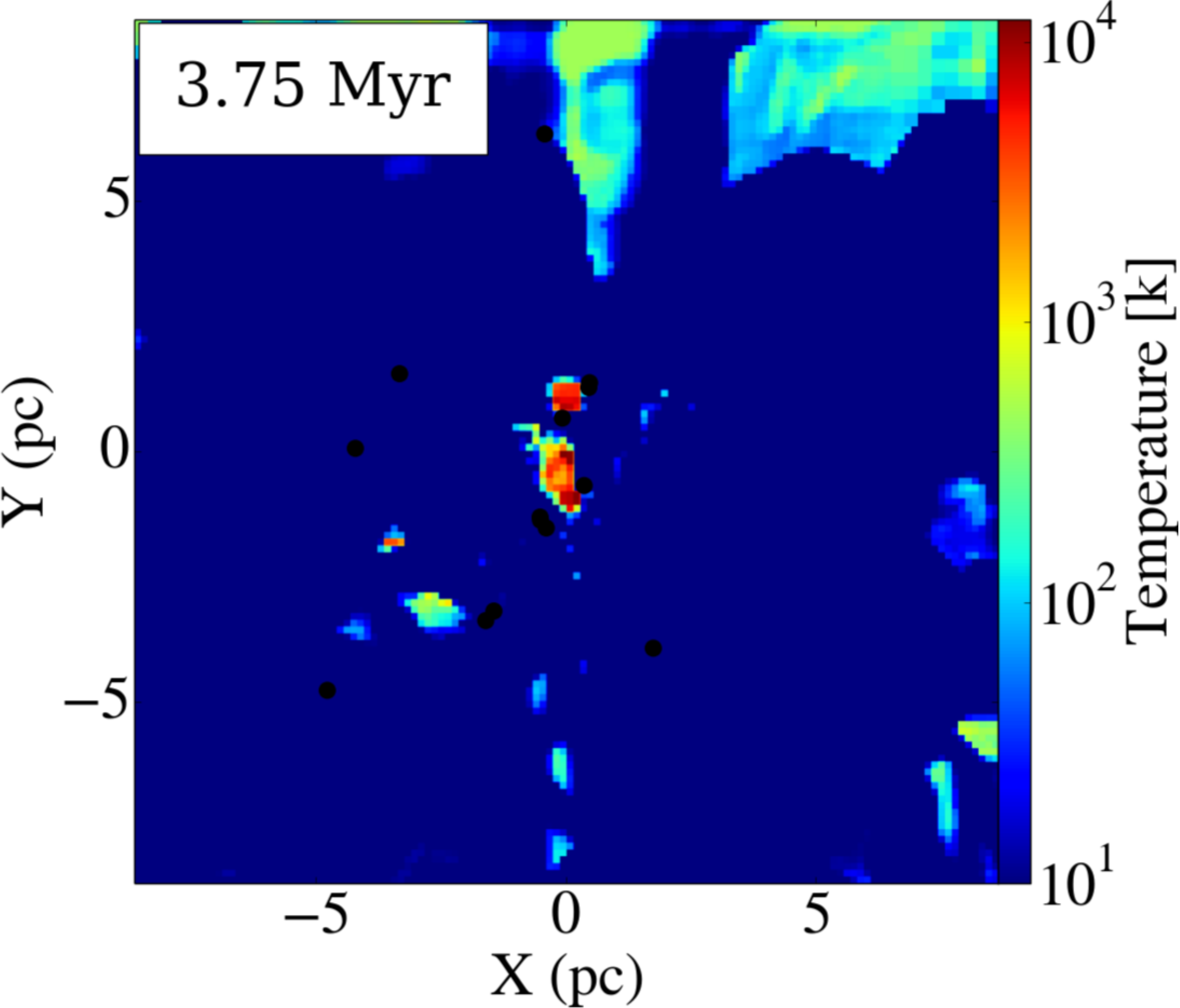} & \includegraphics[width=0.33\linewidth]{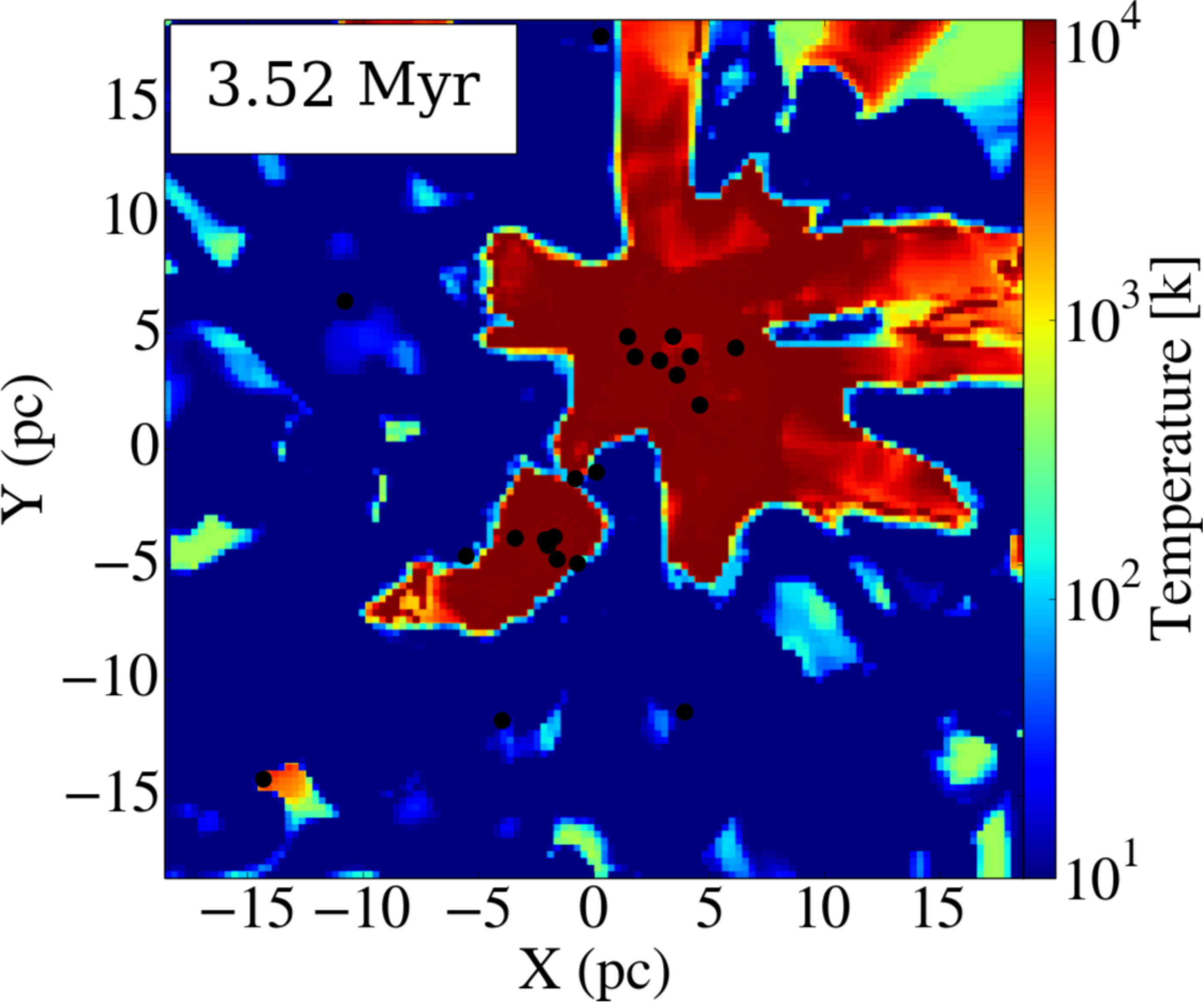} & \includegraphics[width=0.33\linewidth]{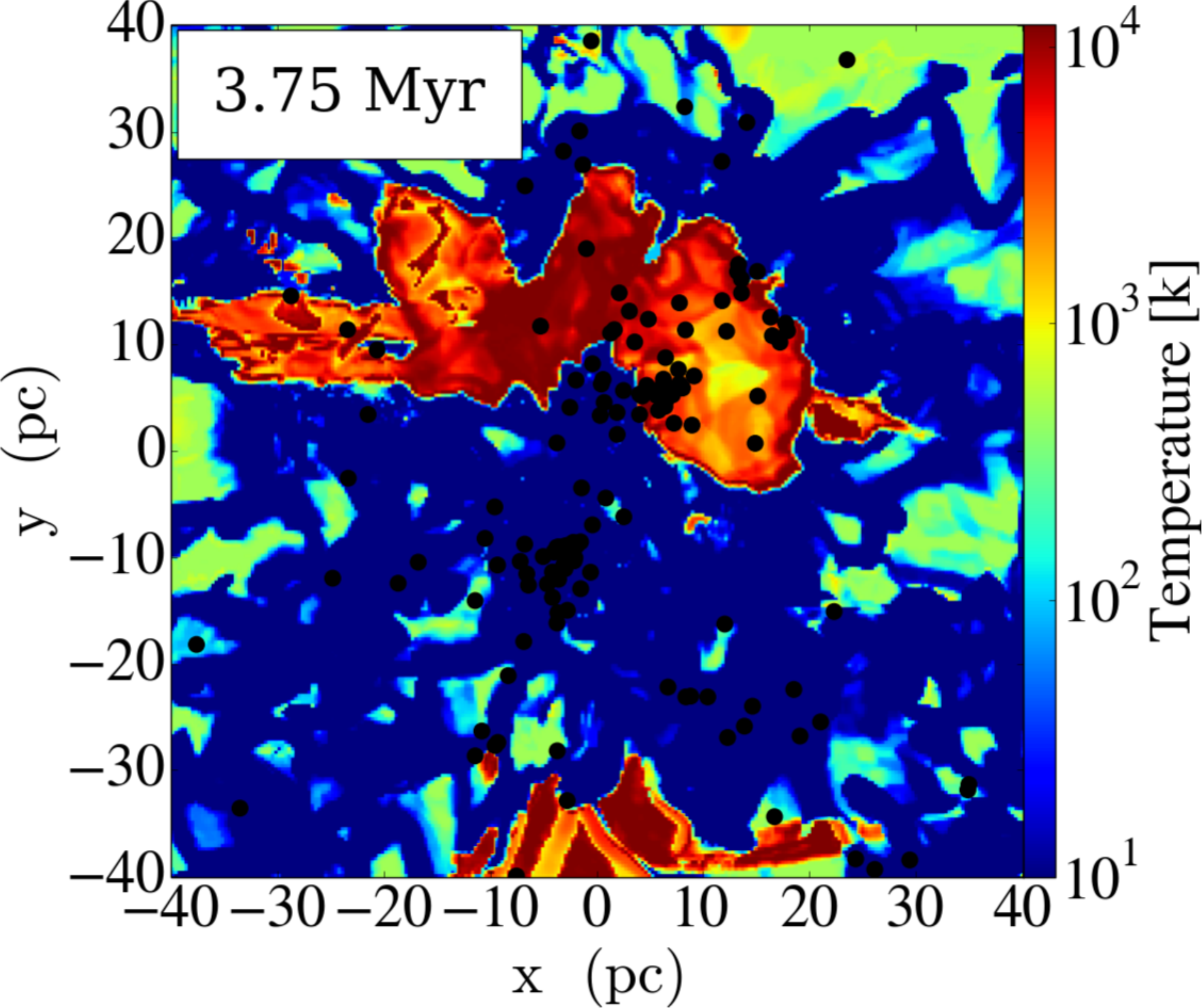} \\
\includegraphics[width=0.33\linewidth]{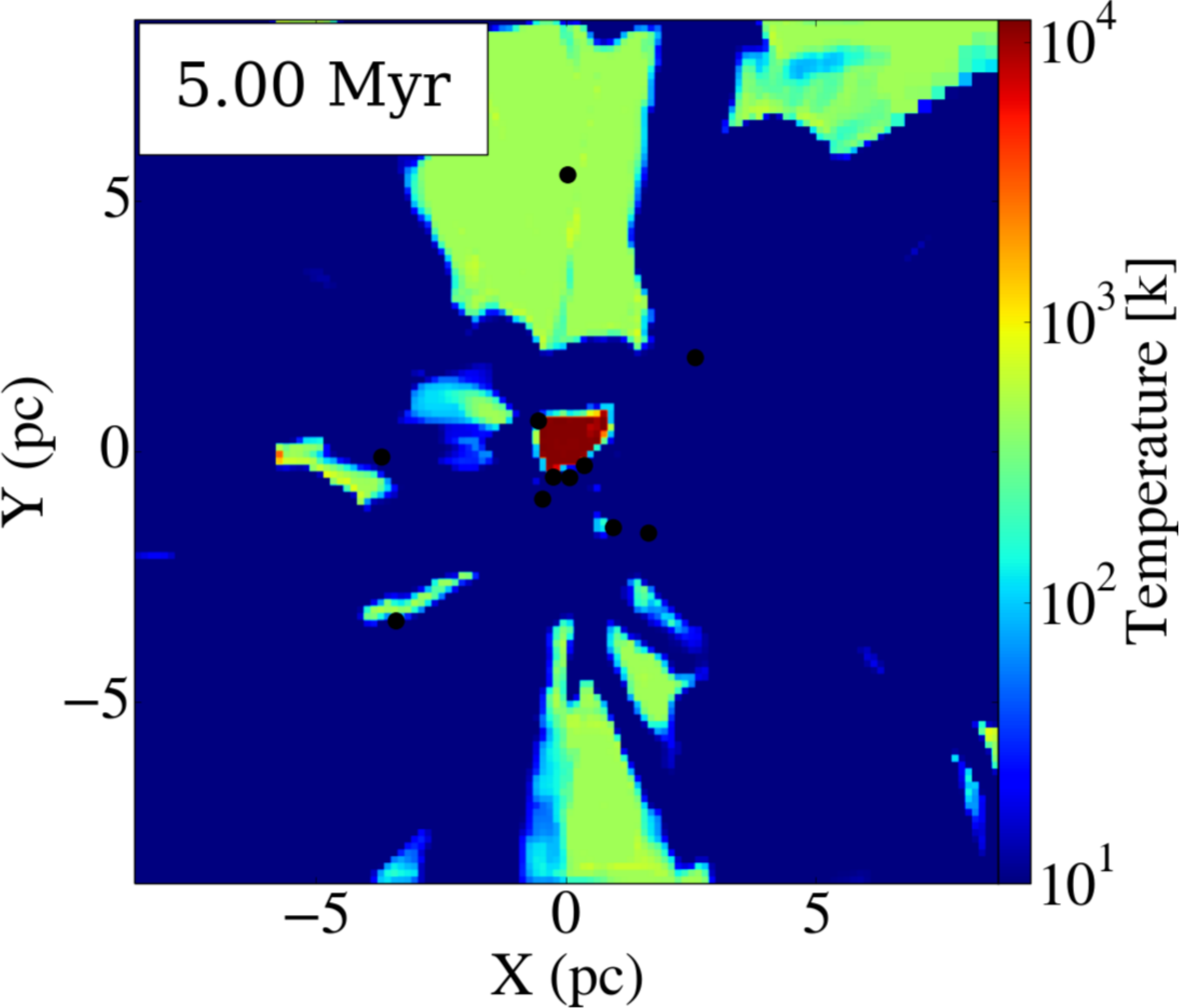} & \includegraphics[width=0.33\linewidth]{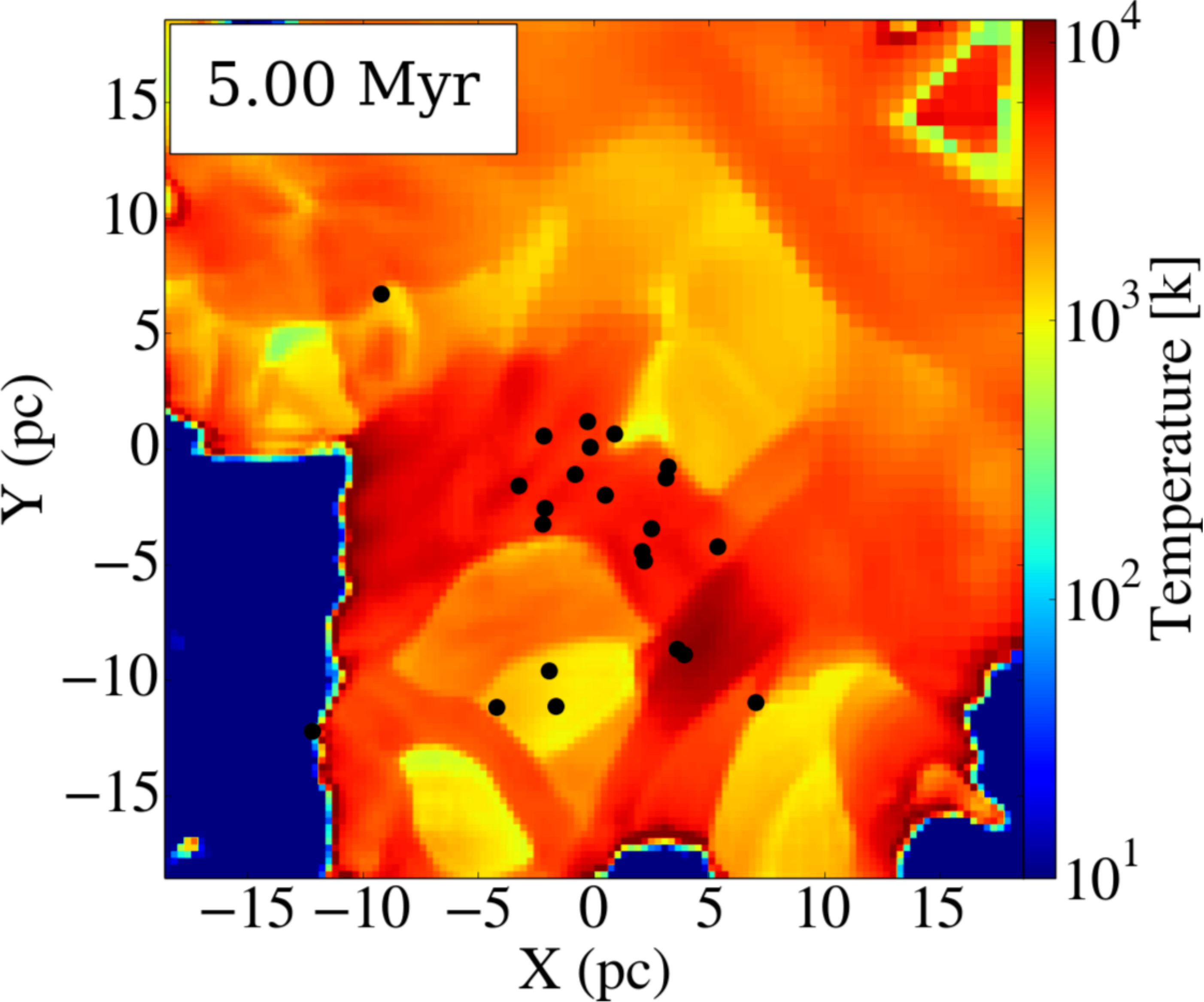} & \includegraphics[width=0.33\linewidth]{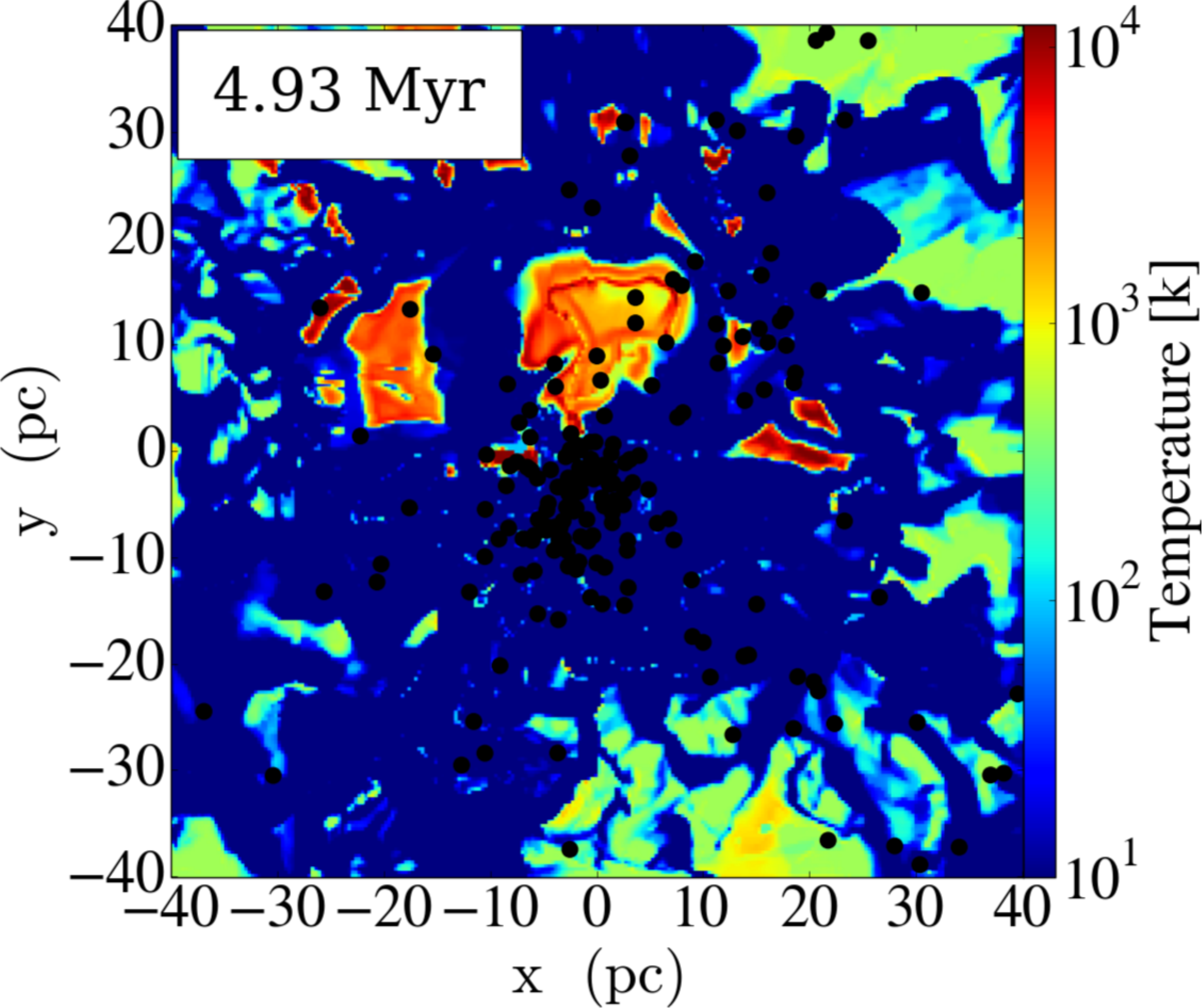} \\
\end{tabular}
\caption{The corresponding temperature slices to the panels shown in Figure 1.}
\label{fig:slice_temp}
\end{figure*}

The hybrid-characteristics raytracing scheme developed by \citet{Rijkhorst}, and expanded for astrophysical use by \citet{Peters2010}, is used to treat radiative transfer.
This scheme follows the propagation of ionizing and non-ionizing radiation and uses the DORIC routines \citep{Frank1994,Mellema2002} to calculate the ionization state of the gas. 
The DORIC routines consider hydrogen to be the only gas species when calculating the absorption of ionizing photons. We adopt the temperature dependent Planck mean opacities from \citet{Pollack1994} for non-ionizing radiation which were calculated for 
a mixture of gas, silicates, ices, and organics. The absorption of non-ionizing radiation acts as a source term when calculating the temperature of the gas.

Radiation pressure is included by adopting a single UV opacity of $\kappa$ = 775 cm$^2$ g$^{-1}$ \citep{LiDraine2001} which is scaled by the neutral fraction of the gas such that fully ionized regions have zero
opacity. The radiative force per unit mass is calculated via,

\begin{equation}
F = \frac{L}{c}\frac{e^{-\tau_{uv}}}{4\pi r^2}
\end{equation}

\noindent where $c$ is the speed of light, $L$ is the source luminosity, $r$ is the distance between the source and the cell, and $\tau_{uv}$ is the optical depth between the source 
and the cell calculated using the raytracer. We note that the scattering and absorption of infrared (IR) photons is not included in our radiation pressure calculation. The 
trapping of IR photons in high density regions would introduce an additional factor of $\tau_{IR}$ --- the optical depth to IR radiation --- in Equation 1. For typical MW cluster forming regions, this additional contribution is thought to be negligible \citep{Murray2010}.    

The formation of star clusters is represented by the sink particle methods from \citet{Federrath2010}. A custom subgrid model is used to model star formation within these clusters 
(see \citet{Howard2014} for a full description of this model). When a particle is formed above the adopted density threshold of 10$^4$ cm$^{-3}$, which is based on observations of
star-forming clumps \citep{Lada2003}, its mass is divided into two components; mass available for star formation during this timestep, and the remaining gas mass (referred to as the 'reservoir').
The mass available for star formation is drawn from the available gas reservoir and randomly distributed into stars using a \citet{Chabrier2005} IMF. The reservoir gas is converted to stars with an efficiency of 20\% per 
freefall time, where the freefall time is taken to be 0.36 Myr (ie. the freefall time of gas at 10$^4$ cm$^{-3}$ with a mean molecular weight of 2.14), and the IMF is sampled every tenth of a freefall time to ensure cluster properties evolve smoothly over time. 
The efficiency per freefall time was chosen to be consistent with observations of star-forming clumps which are estimated to range from 10-30\% \citep{Lada2003}. 

The masses of all stars formed in each cluster are recorded and analytic fits from \citet{Tout1996} are used to obtain each star's temperature from its mass. We neglect the effects 
of protostellar evolution and assume each star to be radiating as a blackbody at its corresponding temperature. The total luminosity of each star is calculated by integrating 
the entire blackbody spectrum and the ionizing luminosity is calculated using the same method but only considering photon energies greater than 13.6 eV. The total ionizing 
luminosity of each cluster is then the sum of its constituent stars which is used by the radiative transfer scheme.

We allow our cluster sink particles to merge under the conditions that they are separated by less than a particle radius, their relative velocities are negative, and they are gravitationally bound to one another.
When a merger occurs, all mass (including both the stellar mass and reservoir mass) is transferred to the more massive particle and the smaller particle is deleted. The total number of 
clusters may therefore either increase or decrease as the simulations evolves. 

We employ a stellar mass threshold for our clusters, below which the clusters do not radiate.
This was included in order to reduce the computational time, since the radiative transfer scheme is expensive. Clusters below this threshold continue to form stars, accrete gas, and interact 
gravitationally with their surroundings, but they are not included in the radiative transfer calculation. We discuss the specific thresholds we used for each simulation below. 

\subsection{Initial GMC Conditions}

\indent \indent We simulate a suite of GMCs that have masses of 10$^4$, 5$\times$10$^4$, 10$^5$, 5$\times$10$^5$, and 10$^6$ M${\odot}$. Two simulations were completed for each cloud mass --- one with radiative feedback included,
and one without radiative feedback (ie. purely hydrodynamic). The clouds are initially spherical, with a density profile which is uniform in the inner half of the cloud and decreases as r$^{-3/2}$
in the outer half. A quadratic fit is applied at the transition between these two profiles to ensure the density is smooth and continuous. The radius of each cloud is chosen
such that the average density is $n$ = 100 cm$^{-3}$. 

Each GMC is initially overlaid with a Burgers turbulent velocity spectrum, as in \citet{2011MNRAS.413.2741G}, after which the turbulence is not driven and allowed to decay. The 
strength of the turbulence varies between simulations but is chosen such that each cloud has the same initial virial parameter, $\alpha_0$, defined by \citep{BertoldiMckee},

\begin{equation}
\alpha_0 = 2\frac{E_{kin}}{|E_{grav}|}
\end{equation}

\noindent where $E_{kin}$ is the cloud's total kinetic energy, and $E_{grav}$ is the total gravitational potential energy. We have chosen an initial virial parameter of 3 (ie. unbound) 
since it resulted in more realistic formation efficiencies compared to bound clouds in \citet{Howard1}. As shown in that work, the turbulence decays rapidly and becomes virialized at $\sim$2.5 Myr regardless of $\alpha_0$.

We use outflow boundary conditions for all simulations. The total mass in the simulation volume is therefore not conserved, and can decrease over time due to gas leaving the domain.
This is relevant to the discussion that follows in the next Section.

Since the radius, initial Mach number, resolution, particle size (given by 2.5 times the smallest cell size), and the threshold for radiating differ between clouds of different mass, we summarize these parameters in Table 1. 

\section{Results}

\subsection{Global Evolution and Cluster Properties}

\begin{figure}
\includegraphics[width=1.\linewidth]{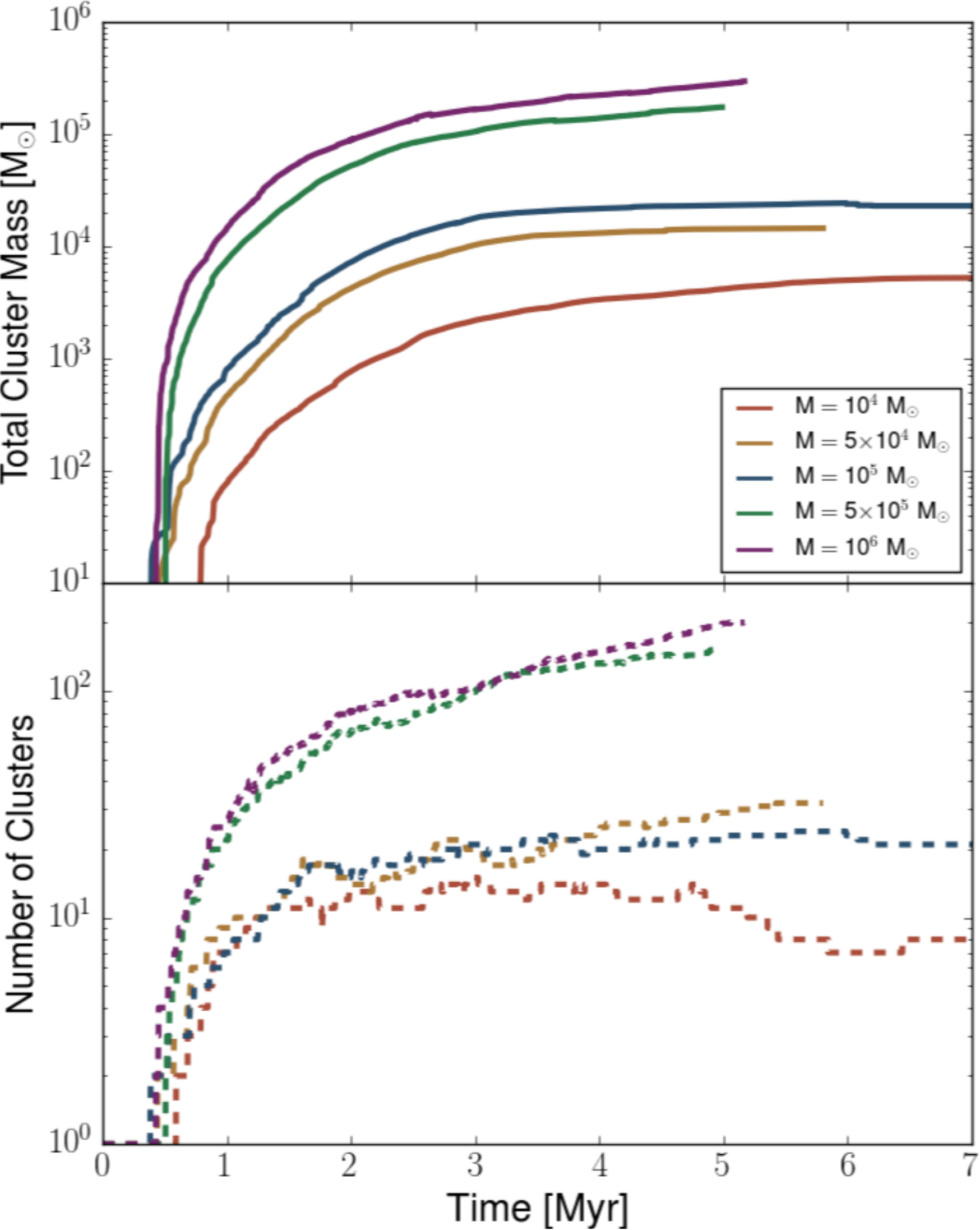} 
\caption{The total mass contained in clusters (top) and the total number of clusters (bottom) in our five GMC models including radiative feedback. Note that the total mass
contained in clusters can decrease over time due to clusters leaving the simulation volume, and the number of clusters can decrease both through escaping clusters and merging events.}
\label{fig:nstars}
\end{figure}

\indent \indent To visually compare the evolution of GMCs with different initial masses, we show density slices through the center of the simulation volumes in Figure \ref{fig:slice_dens}. The columns, from left to right, 
show GMC masses of 10$^4$, 10$^5$, and 10$^6$ M$_{\odot}$ respectively. All simulations shown in Figure 1 include radiative feedback. The rows are plotted at different times, ranging from 1.5 to 5 Myr. The black dots represent the 
locations of clusters which have been projected onto the slice plane. The corresponding temperature slices are shown in Figure \ref{fig:slice_temp}. It is very important to note that cloud sizes and simulation boxes are very different 
for these three GMCs: 10x10 pc, 20x20 pc, and 40x40 pc, respectively. 

The first row shows the state of the simulation shortly after the formation of the first clusters. The gas has already broken up into filaments due to the turbulent nature of 
the gas. The 10$^6$ M$_{\odot}$ simulation has formed significantly more clusters by this time, totaling 37 compared to the 7 that have formed in the 10$^4$ M$_{\odot}$ cloud. 
Despite clusters being present, they have not grown to high enough masses to influence their environment via heating or ionization. This can be seen in the first row of 
Figure 2 which shows that the majority of the gas still remains at 10 K, with $\sim$300 K gas filling the low density voids between filaments.

\begin{figure}
\includegraphics[width=1.0\linewidth]{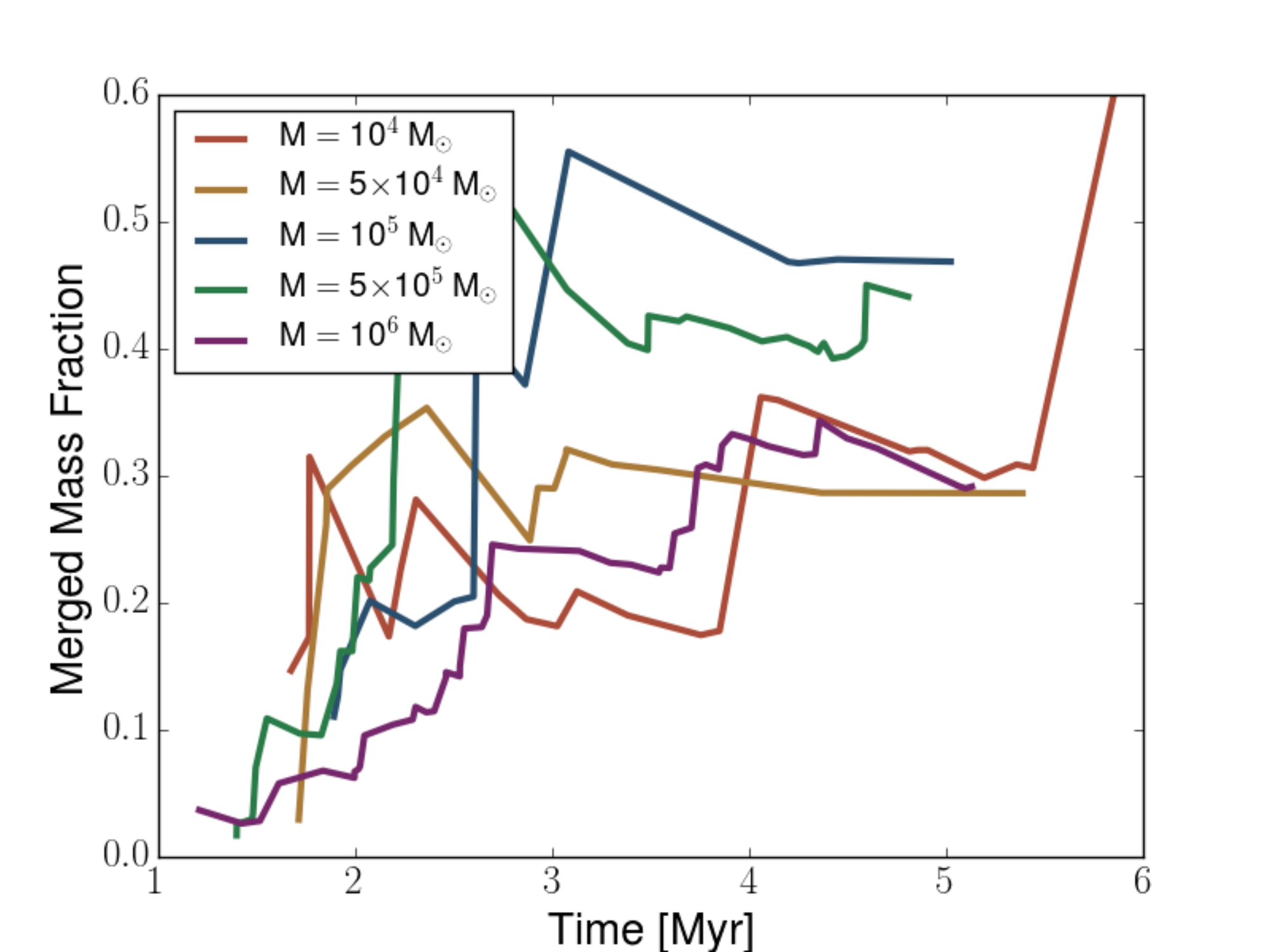} 
\caption{The merged mass fraction, defined as the total cluster mass that has participated in at least one merging event divided by the total mass contained in all clusters, versus time for the 5 GMCs including radiative feedback.}
\label{fig:mergers}
\end{figure}

As the simulation progresses to 2.5 Myr, the clusters in the 10$^6$ M$_{\odot}$ cloud have become sufficiently populated with massive stars to begin ionizing their surroundings. This results in a hot ($\sim$10,000 K)
bubble of gas near the center of the simulation shown in Figure \ref{fig:slice_temp}. The corresponding density slice shows that filaments in this region have been destroyed due to the high temperatures. The 10$^4$ and 10$^5$ M$_{\odot}$ clouds, in
contrast, have not produced enough massive stars for radiative feedback to have any effects.

\begin{figure*}
\includegraphics[width=1.0\linewidth]{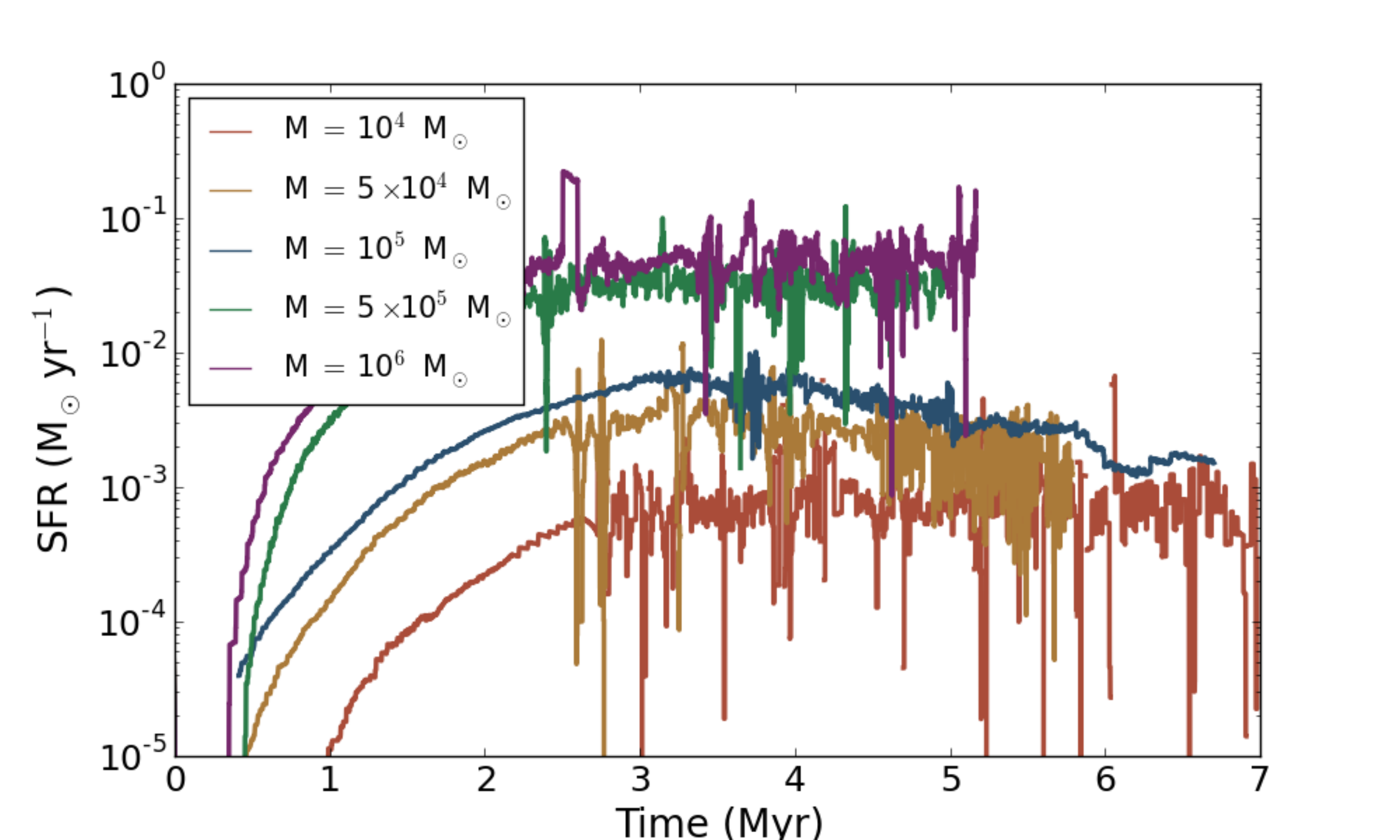} 
\caption{The total star formation rates (SFRs) for the 5 simulated GMCs which include radiative feedback. Note that this plot has been smoothed for readability (see text).}
\label{fig:SFR}
\end{figure*}

At 3.75 Myr, Figure \ref{fig:slice_temp} clearly shows that radiative feedback is active in all clouds. The 10$^5$ M$_{\odot}$ and 10$^6$ M$_{\odot}$ clouds in particular show extended HII regions centered on
a group of massive clusters. The corresponding density images show that radiative feedback is in the process of destroying the filaments in the vicinity of these HII regions due
to the expansion of the hot gas which smears out overdense regions. 

The final panels of Figures \ref{fig:slice_dens} and \ref{fig:slice_temp} show marked differences between the three simulations. The gas in the 10$^4$ M$_{\odot}$ cloud is centrally condensed with the majority of clusters existing in
this central region. This allows these clusters to continue accreting from their surroundings. 

The 10$^5$ M$_{\odot}$ cloud has been effectively destroyed by radiative feedback. The entire
cloud is nearly fully ionized and the resulting expansion of gas has caused a large fraction of the initial mass to leave the simulation volume. The cloud remains fully ionized
after this point and the accretion of gas by the clusters has been halted.

While there are large voids produced by HII regions in the 10$^6$ M$_{\odot}$ cloud, little mass loss has occurred. The clusters are also dispersed more evenly throughout the cloud,
some of which are still actively accreting gas. Large scale filamentary structures are still present at 5 Myr. 

To compare cluster formation across different clouds, we plot the total mass contained in clusters (top panel) and the total number of clusters (bottom panel) in Figure \ref{fig:nstars}.

\begin{figure*}
\includegraphics[width=0.98\linewidth]{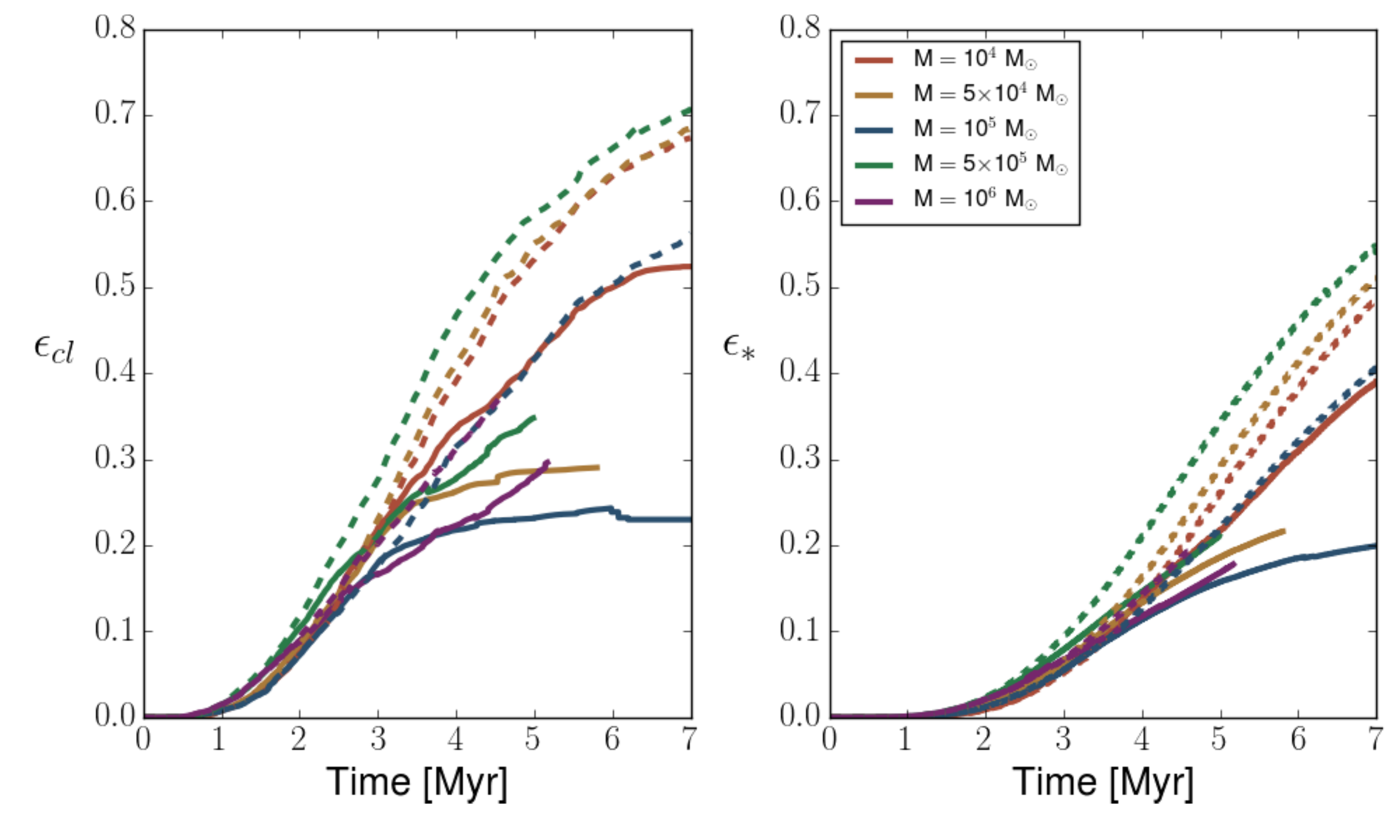} 
\caption{Left: The cluster particle formation efficiency ($\epsilon_{cl}$, defined as the total mass in cluster particles divided by the GMC's initial mass) for our 5 RHD simulations (shown by
solid lines) and the 5 HD simulations (shown by dashed lines). Right: Identical to the left panel except the star formation efficiency ($\epsilon_{*}$, total mass of stars within clusters
divided by the initial GMC mass) is plotted.}
\label{fig:eff}
\end{figure*}

Since all GMCs were initialized with the same average density and virial parameter, the onset of cluster formation is comparable, ranging from 0.39 Myr for the 10$^5$ M$_{\odot}$ cloud to 0.59 Myr at 10$^4$ M$_{\odot}$. The clusters then rapidly
grow in mass via gas accretion with the higher mass clouds containing more mass in clusters at any given time, as expected. At 5 Myr, the total mass contained in clusters, in order of lowest to highest initial cloud mass, 
is 4.1$\times$10$^3$, 2.3$\times$10$^4$, 2.3$\times$10$^4$, 1.8$\times$10$^5$, and 2.8$\times$10$^5$ M$_{\odot}$. Note that the total cluster mass does not scale directly with the initial cloud mass. This will be relevant to the discussion of formation efficiencies which follows.

The number of clusters formed also does not scale directly with the initial cloud mass. The numbers of clusters at 5 Myr in the 10$^4$, 10$^5$, and 10$^6$ M$_{\odot}$ simulations are 11, 22, and 199, respectively. Note that we allow our clusters to merge so
the number of clusters can decrease. Since mass is conserved in the merger, however, the total mass in clusters cannot decrease unless a cluster leaves the simulation volume entirely which does not play a significant role. Only the 5$\times$10$^5$ M$_{\odot}$ and 10$^6$ M$_{\odot}$ GMCs are
still forming clusters in significant numbers at the end of the simulation.

Cluster merging plays a significant role in the growth of clusters. We demonstrate this in Figure \ref{fig:mergers} which plots the merged mass fraction versus time. We define the merged mass fraction as total amount of cluster mass that has participated in a merger event up until a given time divided by the total mass contained in clusters at that same time. A merged mass fraction of 0.5, for example, means that half of the mass contained in clusters has participated in at least one merger.

While there does not appear to be a trend with GMC mass, it is clear that significant numbers of mergers are occurring in all clouds. At $\sim$5 Myr, the merged mass fractions range from 0.28 to 0.47. A fraction of 0.28 indicates that cluster
growth is dominated by gas accretion while a fraction of 0.47 represents comparable contributions from gas accretion and cluster mergers. This highlights the importance of cluster merging during the early phases of star formation and suggests that a 
combination of both gravitational fragmentation (ie. top down cluster formation) and hierarchical merging (ie. bottom up cluster formation) are needed to fully understand the formation and evolution of young stellar clusters.

Lastly, we compare the star formation rates (SFR) in our GMC models in Figure \ref{fig:SFR}. Note that this refers to the formation of stars \textit{within} the clusters and not the formation of new clusters.

As a product of our star formation subgrid model which samples the IMF to form new stars at prescribed intervals, there are timesteps in which no new stars formed and others 
which have a burst of star formation. We have therefore smoothed these plots using a sliding average window to assist in readability. This leads to a highly variable SFR, particularly
at late times when there are many clusters forming stars at staggered times.

All curves show a sharp rise in SFR following the onset of cluster formation. For the 10$^4$, 5$\times$10$^5$, and 10$^6$ M$_{\odot}$ GMCs, the SFR levels out to approximately constant 
values of 6$\times$10$^{-4}$, 3$\times$10$^{-2}$, and 5$\times$10$^{-2}$ respectively. These values are consistent with a SFR that scales directly with the initial GMC mass, assuming
similar density structures, as found in observations of local GMCs by \citet{Lombardi2010}. 

The other two GMC models (5$\times$10$^4$ and 10$^5$ M$_{\odot}$) instead show a SFR rate which decreases at late times. As shown in \citet{Howard2014}, which examined the properties
of our adopted subgrid model for star formation, a decreasing SFR is indicative of a population of clusters which have completely stopped accreting and are simply using up the 
rest of their gaseous reservoir. The images for the 10$^5$ M$_{\odot}$ GMC in Figures \ref{fig:slice_dens} and \ref{fig:slice_temp} are consistent with this picture since they demonstrate that the cloud has been almost fully ionized and destroyed by 5 Myr. This suggests the impact radiative feedback has on the formation and evolution of clusters is stronger in these clouds. We compare the effects of radiative feedback between cloud models below.

\subsection{The role of radiative feedback}

To understand the role of radiative feedback in GMCs with different initial masses, we computed a grid of complementary simulations which have radiative transfer turned off. We will refer to simulations with radiative feedback included as
"RHD" (Radiation Hydrodynamics) simulations and "HD" (Hydrodynamics) simulations are those with radiative feedback not included. 

How much the efficiency is suppressed when including radiative feedback is
still debated. \citet{Howard1} showed that it depends on the initial gravitational boundedness of the molecular cloud, as measured by the virial parameter. Here, we find that radiative feedback does indeed limit star cluster formation but, more importantly, the strength of this suppression depends on the cloud's initial mass.

We show this in Figure \ref{fig:eff}, which plots the cluster particle formation efficiency ($\epsilon_{cl}$) and the star formation efficiency ($\epsilon_{*}$) for both the RHD simulations, shown by the solid lines, and the HD simulations, shown by the dashed lines. We define 
$\epsilon_{cl}$ as the total mass in cluster particles divided by the initial cloud mass. The star formation efficiency, $\epsilon_{*}$, is defined as the total mass of stars within the clusters divided by the initial cloud mass. Note that an entire cluster's mass
is not necessarily only in stars, but can also be part of the gas reservoir which is available for future star formation.

When comparing the RHD and HD simulations, we see that star and cluster formation start at similar times and evolve identically for $\sim$2.5 Myr. At this point, $\epsilon_{cl}$ in the HD and corresponding RHD runs begin
to diverge, with the HD simulations having the higher efficiency in all cases. This trend continues to the end of the simulation and the difference between the HD and RHD formation efficiencies grows. Choosing a time of 5 Myr to compare $\epsilon_{cl}$, the efficiencies
in the lowest to highest mass clouds are 43\%, 29\%, 23\%, 35\%, and 28\%. At the same time, $\epsilon_{*}$ ranges from 16\% to 21\%. Note that these values are higher than the measured values from GMC observations (eg. \citet{Lada2003, McKee2007,Murray2011}) which 
suggests that while radiative feedback does lower $\epsilon_{*}$ relative to HD runs, other pieces of physics such as stellar winds are required to lower these values further. 

Both $\epsilon_{cl}$ and $\epsilon_{*}$ are the highest for the lowest mass (10$^4$ M$_{\odot}$) GMC. This is consistent with the results from \citet{Ochsendorf} who 
found evidence of a decreasing $\epsilon_{*}$ with increasing cloud mass in the LMC. While we reproduce their results for the 10$^4$ M$_{\odot}$ cloud, we do not see clear 
evidence for a trend with increasing GMC mass.

\begin{figure*}
\includegraphics[width=1.0\linewidth]{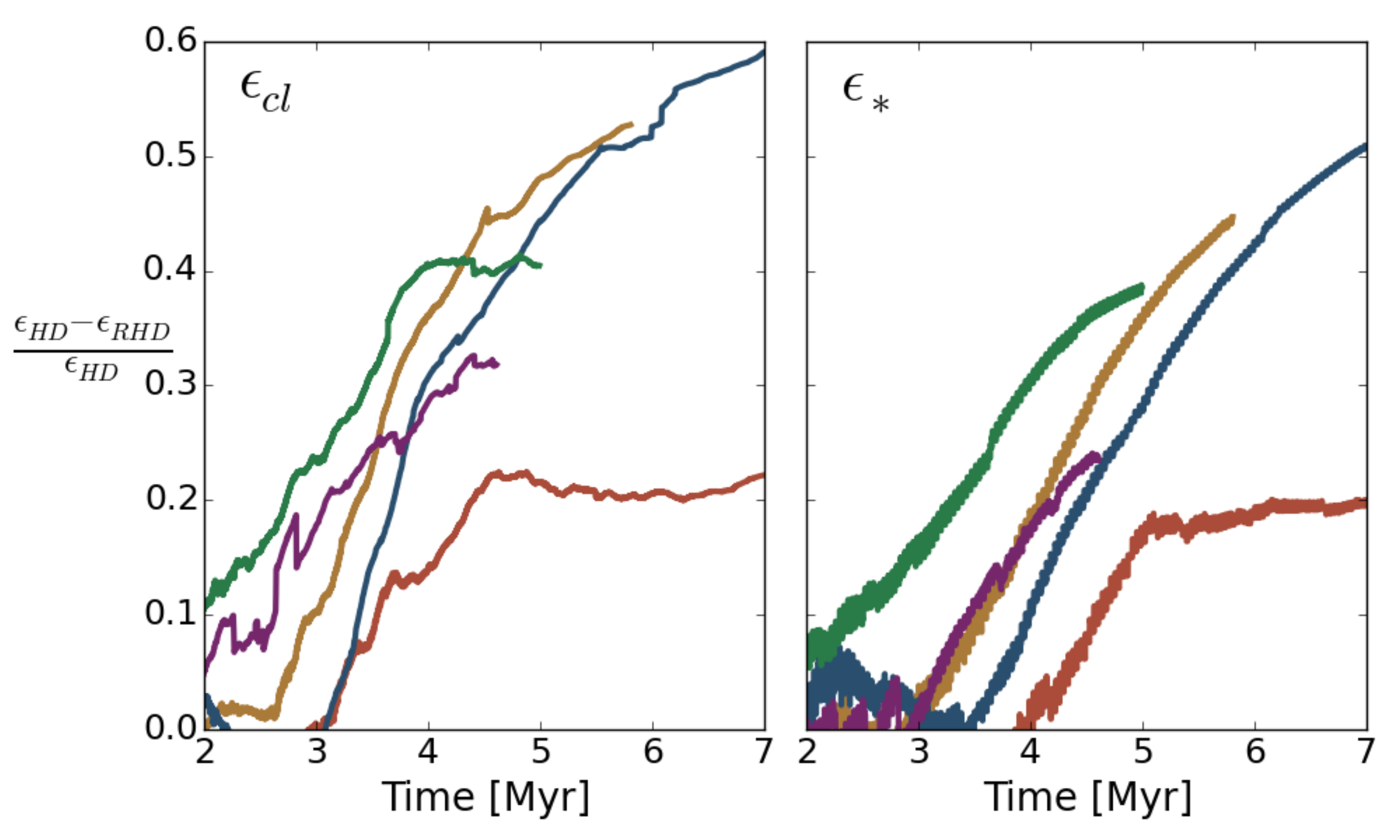} 
\caption{The fractional reduction of $\epsilon_{cl}$ (left) and $\epsilon_{*}$ (right) when including radiative feedback into a simulation, relative the HD formation efficiencies.}
\label{fig:diff}
\end{figure*}

It is clear from Figure \ref{fig:eff} that radiative feedback plays a stronger role in suppressing star and cluster formation in some clouds more than others. To make this clear,
we plot the fractional reduction in efficiencies when including radiative feedback in Figure \ref{fig:diff}. This Figure shows that difference in formation efficiencies is 
largest for the 5$\times$10$^4$ and 10$^5$ M$_{\odot}$ GMCs. Focusing on these two simulations at 5 Myr, the difference between $\epsilon_{cl}$ for the HD and RHD run is 27\% and 
18\% for initial masses of 5$\times$10$^4$ and 10$^5$ M$_{\odot}$ GMCs, respectively. This corresponds to approximately a factor of two reduction in both cases. The inclusion 
of radiative feedback in the 10$^4$, 5$\times$10$^5$, and 10$^6$ M$_{\odot}$ GMCs reduced $\epsilon_{cl}$ by 21\%, 40\%, and 33\% relative to the HD simulations, respectively. 

This is consistent with the density and temperature visualizations discussed in Figures \ref{fig:slice_dens} and \ref{fig:slice_temp}. It was clear from those images that the 10$^5$ M$_{\odot}$ simulation is
more globally impacted by radiative feedback than the other two cases, as evidenced by the nearly fully ionized simulation volume. In contrast, the 10$^4$ and 10$^6$ M$_{\odot}$ GMCs showed small HII 
regions which may stop the accretion onto local clusters but not the entire population.

\begin{figure}
\includegraphics[width=1.0\linewidth]{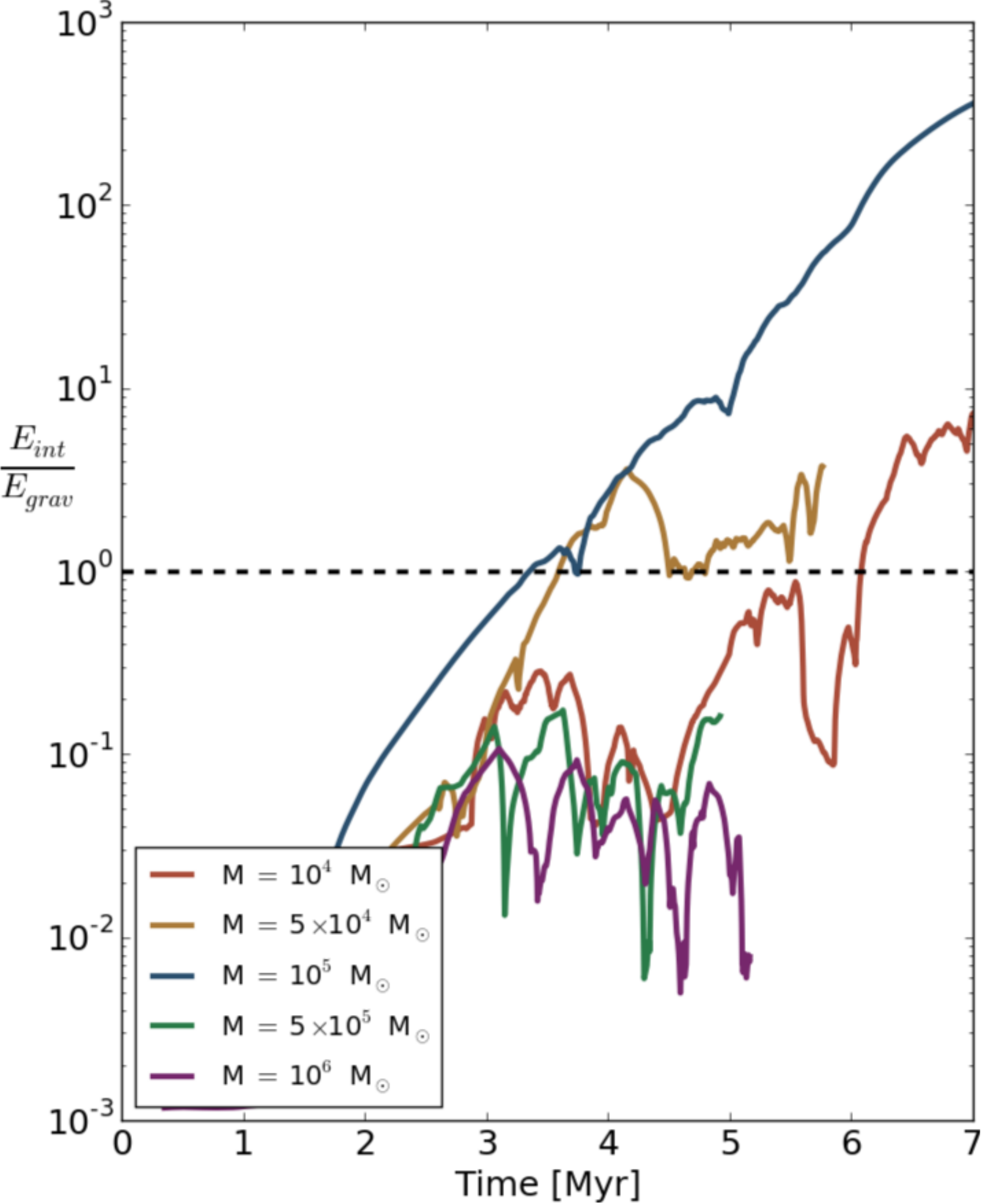} 
\caption{The ratio between the total internal energy of the gas to the global gravitational potential energy of the cloud. A lower ratio suggests a higher suppression of cluster and star formation.}
\label{fig:ratio}
\end{figure}

These results suggest there is something unique happening in clouds between 5$\times$10$^4$ - 10$^5$ M$_{\odot}$ which make them more susceptible to radiative feedback effects. 
We propose that clouds lower than this mass are not able to form enough massive stars and therefore cannot completely ionize the cloud. Indeed, the 10$^4$ M$_{\odot}$ did not 
produce any O-stars throughout its evolution. On the other hand, clouds above this mass range are capable of producing O-stars but have too much gas mass, and therefore a higher column density
to ionizing radiation, to be fully ionized during the early stages of cluster formation. The overall gravitational potential of these massive clouds is also deeper, meaning they are harder
to unbind overall.

We can demonstrate this balance between gravity and the energy injected by radiative feedback by comparing the total gravitational potential energy to the total internal energy of 
the gas at any given time. The internal energy is calculated from the gas temperature and ionization fraction and is therefore a proxy for the energy injected by radiation. We
plot the ratio of the total internal energy to the total gravitational potential energy in Figure \ref{fig:ratio}.

\begin{figure}
\includegraphics[width=1.0\linewidth]{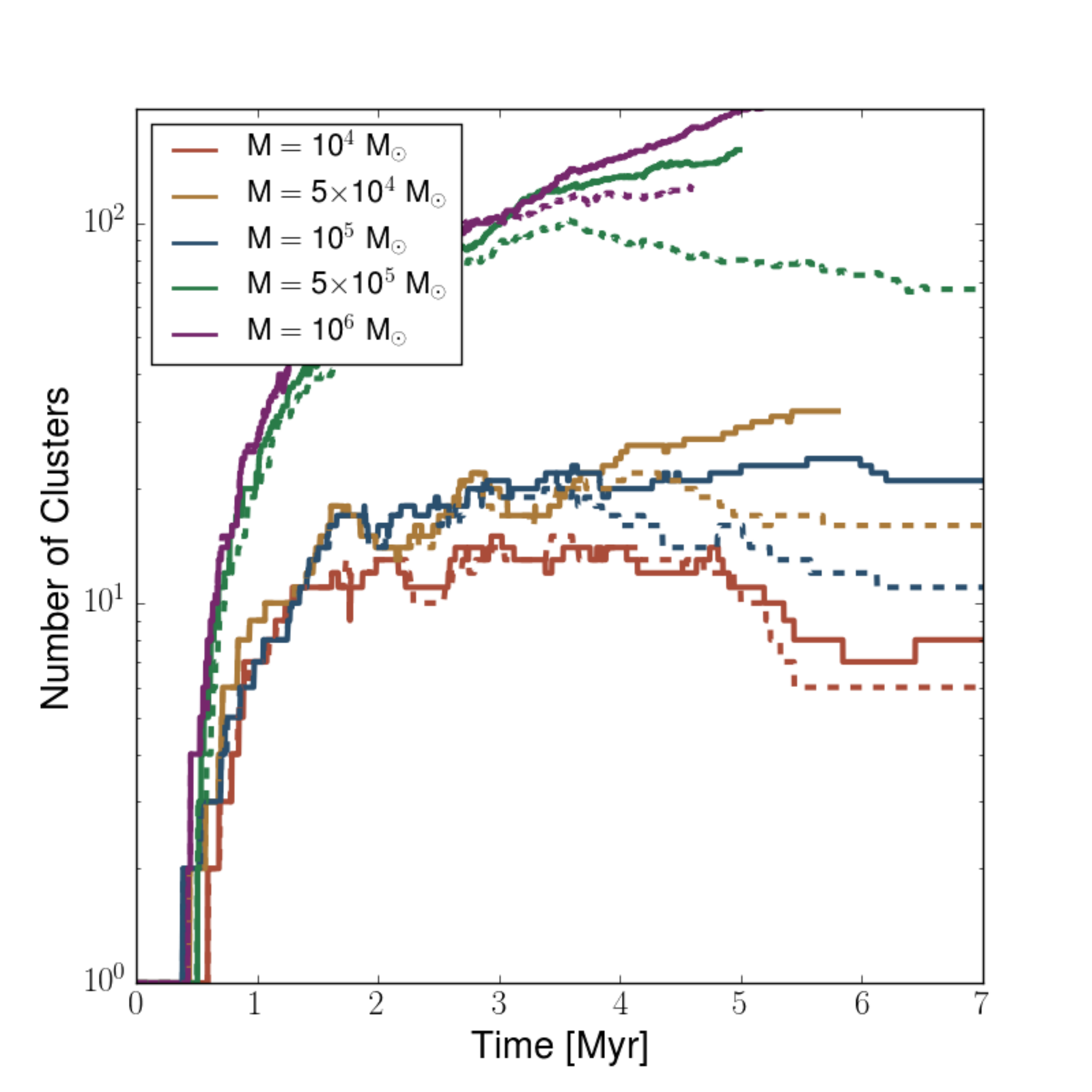} 
\caption{The number of clusters as a function of time (as seen in Figure \ref{fig:nstars}) including both RHD (solid) and HD (dashed) simulations.}
\label{fig:nstars2}
\end{figure}

All models start with a ratio less than 1, indicating that gravitational potential energy dominates during early times. At approximately 3.5 and 3.8 Myr, the ratio
rises above one for the 10$^5$ M$_{\odot}$ and 5$\times$10$^4$ GMCs, respectively. This indicates that the amount of radiation being injected into the gas is sufficient to
unbind the cloud globally, resulting in the larger suppression of cluster formation under the influence of radiative feedback in these clouds. While internal energy does dominate over gravitational potential
energy for the 10$^4$ M$_{\odot}$ GMC, it only does so at late times and therefore does not influence the early stages of cluster formation as significantly. In contrast, the more massive 
models (5$\times$10$^5$ and 10$^6$ M$_{\odot}$) are always dominated by gravitational potential energy.

These results explain why, in Figure \ref{fig:eff}, $\epsilon_{cl}$ begins to flatten around 3 Myr for the 5$\times$10$^4$ and 10$^5$ M$_{\odot}$ clouds. As a larger volume of gas becomes hot 
and ionized, the formation of new clusters, and the accretion onto existing clusters, is suppressed. A similar result is not seen $\epsilon_{*}$ due to our subgrid model. Since 
we do not allow unused gas to leave the clusters, star formation can proceed regardless of whether accretion is still taking place. As shown in \citet{Howard2014}, cluster masses 
are typically dominated by the reservoir of gas, especially at early times.  

The varying strength of radiative feedback may have important implications for the growth and evolution of GMCs if we assume they form through a bottom up process, such as spiral arm induced collisions, as suggested
by \citet{Dobbs2013}. Our results indicate that once a cloud reaches $\sim$5$\times$10$^4$ M$_{\odot}$, it should be destroyed via radiative feedback. This may act as a bottleneck
for the growth of GMCs and could be partly responsible for their observed mass distribution.

The cluster formation efficiency is essentially a normalized measure of how much mass is present in clusters at any given time. To understand how this mass is distributed, we also
need to know the total number of clusters. We examine how the number of clusters is affected by radiative feedback in Figure \ref{fig:nstars2}. This is similar to the plot shown in the previous
section, except the results from the HD simulations are included as dashed lines. We see that the early evolution of the RHD and HD simulations are similar, but at late times there are 
more clusters present in the RHD cases. This will impact the distribution of cluster masses. Since the HD simulations have more mass contained in clusters (as illustrated in Figure \ref{fig:eff})
but fewer clusters overall, the average cluster mass will be higher than cases which include radiative feedback. Taking, for example, the 5$\times$10$^4$ M$_{\odot}$ cloud, the
final average cluster mass is decreased from 2046 M$_{\odot}$ to 468 M$_{\odot}$ when including radiative feedback. 

\section{Observational Comparisons}

\subsection{The Initial Cluster Mass Function}

The mass function of star clusters have been characterized observationally. As discussed in \citet{FallChandar}, the mass function for embedded clusters \citep{Lada2003} and extragalactic clusters taken from the 
Magellanic Clouds, M83, M51, and Antennae are all consistent with a powerlaw mass distribution of the form dlog(N) $\propto$ M$^{\beta}$dlog(m) where $\beta\sim$-1. Here, we compare the mass functions of our simulated clusters to
these results.

\begin{figure}
\includegraphics[width=1.0\linewidth]{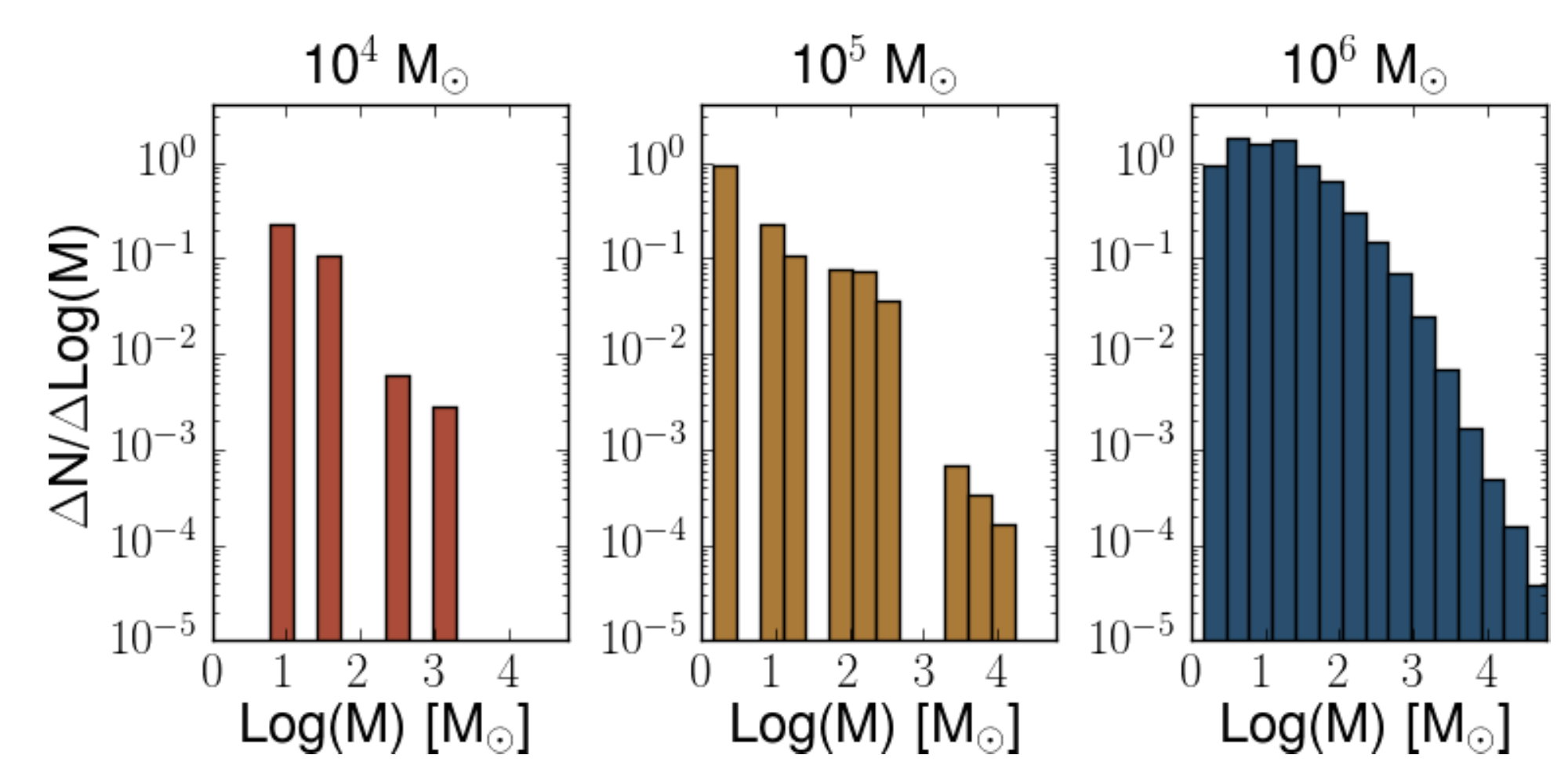} 
\caption{The cluster mass function for the 10$^4$ (left), 10$^5$ (center), 10$^6$ (right) M$_{\odot}$ GMCs, plotted at 5 Myr.}
\label{fig:ECMF_single}
\end{figure}

The cluster mass functions for the 10$^4$, 10$^5$, 10$^6$ M$_{\odot}$ GMCs are shown in Figure \ref{fig:ECMF_single}. The data is plotted at 5 Myr, corresponding to the approximate end of the 10$^6$ M$_{\odot}$ simulation. The mass values represent only the \textit{stellar} mass
contained in each cluster and therefore do not include the unused gas reservoir.

As the initial mass of the GMC increases, the total number of clusters formed also increases (see Figure \ref{fig:nstars}) and so the plot is consequently more populated. 
The cluster mass distributions also shift to higher masses as the initial GMC mass increases. 

\subsection{Cloud Mass - Maximum Cluster mass Relation}

The previous results suggest a relation between the maximum mass cluster produced in a star-forming event
and the mass of the GMC out of which it forms. This relation has been found in both observations \citep{Hughes} and simulations \citep{Fujii} to take the form M$_{c,max}\propto$M$_{GMC}^{0.5}$, 
where M$_{c,max}$ is the maximum mass cluster that forms out of a GMC of mass M$_{GMC}$.

\begin{figure}
\includegraphics[width=1.0\linewidth]{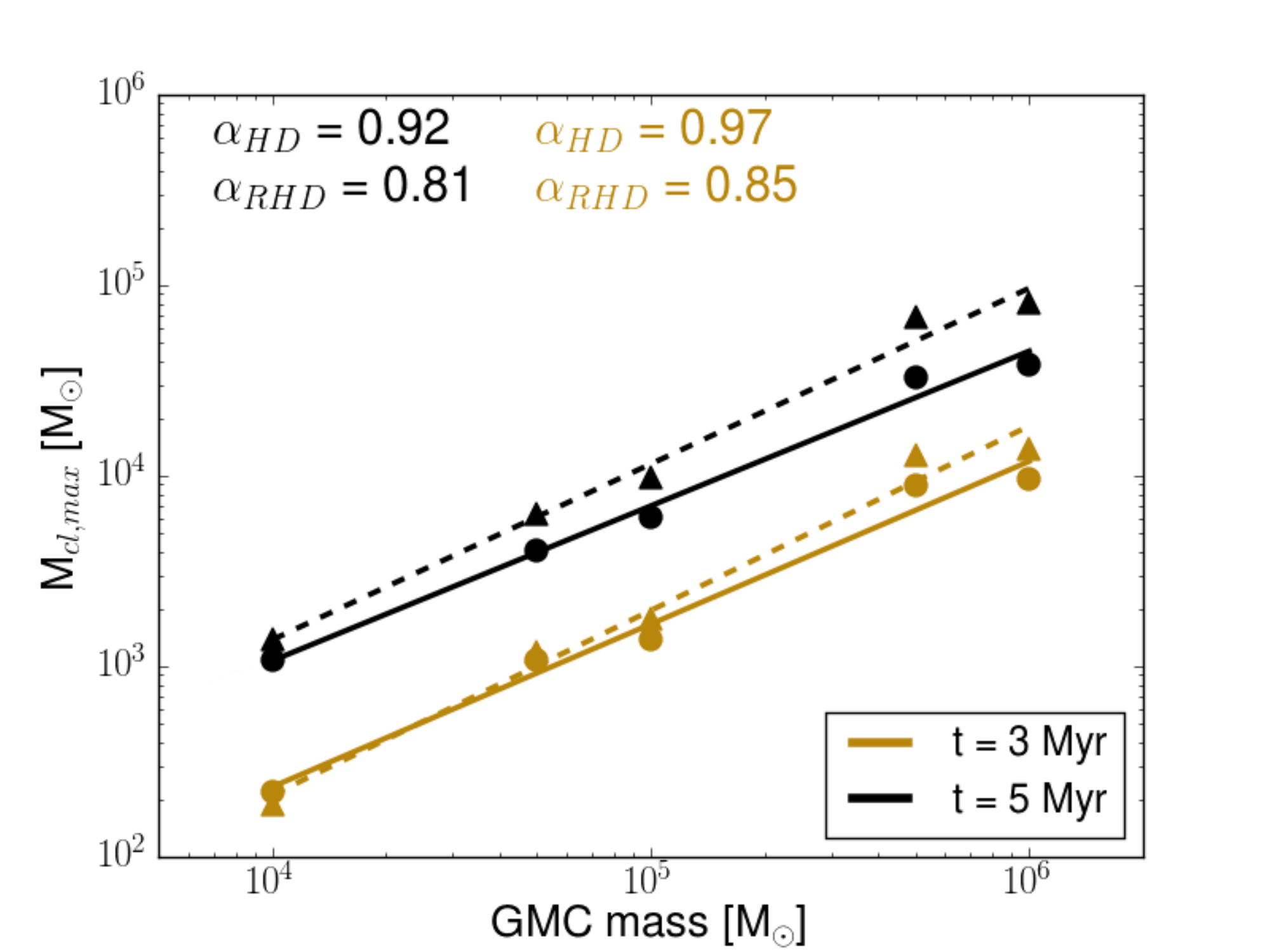} 
\caption{The maximum mass cluster produced in our 5 RHD models (circles) and the 5 HD simulations (triangles) as a function of the initial cloud mass. The results are plotted 
at 3 Myr (gold) and 5 Myr (black). The slope of the distributions are shown at the top of the plot and are colored based on the times they represent. Fits to the HD data are 
shown by the dashed lines, and fits to the RHD data are shown by solid lines.}
\label{fig:maxcl}
\end{figure}

We plot the maximum cluster mass obtained from our 5 RHD GMC models in Figure \ref{fig:maxcl}, shown by the filled circles. We plot the relation at two times --- 3 Myr (gold) and 5 Myr (black). Here, the maximum cluster mass includes only its stellar mass and not the unused reservoir of gas. 

At 5 Myr, we find a relation between the maximum cluster mass and the host GMC mass given by,

\begin{equation}
M_{c,max} \propto M_{GMC}^{0.81 \pm 0.09}.
\end{equation}

\noindent While this does not agree with the relation above, it is roughly
consistent with the relation between the maximum mass star formed in a given cluster proposed by observations from \citet{Pflamm} and the theoretical model of \citet{Elmegreen2002} that have powerlaw indices of
0.67 and 0.74, respectively. This seems to suggest that there may be self-similar star formation processes ranging from GMC masses down to protostellar core masses. In order to 
verify this claim, a fully consistent simulation of a GMC which resolves the formation of individual stars, in combination with a cluster finding algorithm, would be required.

Using the HD simulations at 5 Myr instead, we find a steeper slope of 0.93 and all points are higher than their RHD counterparts. Radiative feedback is clearly limiting the growth
of the most massive clusters regardless of initial GMC mass. The separation between the HD and RHD maximum cluster masses is, however, more pronounced for the larger GMCs. This is likely due
to the large population of massive stars in these clusters that can more effectively heat and ionize their surroundings and suppress further gas accretion.

Figure \ref{fig:maxcl} shows that the slope of these distributions hardly vary with time. At 3 Myr, the slopes of the RHD and HD simulation are 
0.85 and 0.97 (compared to 0.81 and 0.92 at 5 Myr), respectively. The intercept, however, does change from 3 to 5 Myr due to the growth of the stellar populations in these clusters. 
The separation between the HD and RHD clusters is also less pronounced compared to 5 Myr because radiative feedback has not been active for as long.

The insensitivity of the slope with time is likely due to two reasons. Firstly, we are plotting the clusters with the largest stellar content and therefore the highest 
luminosity. For the RHD simulations, these clusters significant affect their local surroundings and suppress their own growth in similar ways. Secondly, our 
subgrid model for star formation prescribes the rate at which stars form in the clusters. Once the most massive clusters accrete a significant amount of gas, the total stellar 
mass will increase at the same rate.

We can estimate the mass of the host GMCs out of which Globular Clusters (GCs) ought to form by extrapolating our relation to larger cluster masses. We note that we have not yet completed 
any GMC simulations greater than 10$^6$ M$_{\odot}$ and, as shown in this work, the effects of radiative feedback are a clear function of cloud mass. It is therefore possible that
the relation displayed in Figure \ref{fig:maxcl} does not extend to higher masses. Assuming it does, a GMC of $\sim$4.5$\times$10$^{7}$ M$_{\odot}$ is required in order to form a GC of mass 10$^6$ M$_{\odot}$.
This is consistent with \citet{HP1994} who argued that Supergiant molecular clouds ($>$10$^{7}$ M$_{\odot}$) are the hosts
to GC formation.  

\subsection{Combined Cluster Mass Function}

\begin{figure}
\includegraphics[width=1.0\linewidth]{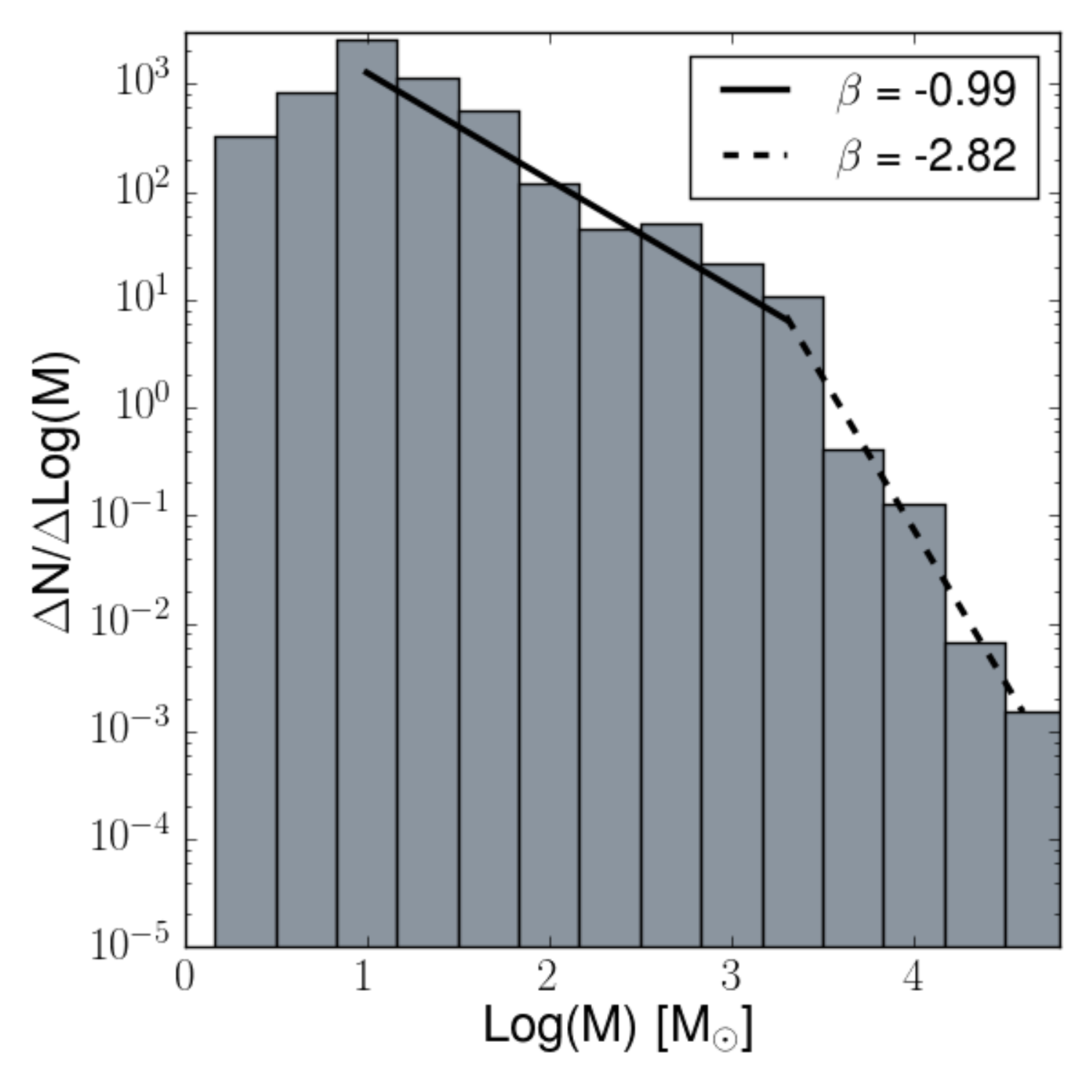} 
\caption{The combined cluster mass function obtained from all 5 RHD simulations. The relative abundance of each parent cloud has been included via the GMC mass
function in order to compare directly to observations. Fits to the mass range of observed embedded clusters (solid) and the high mass regime (dashed) are also included.}
\label{fig:ECMF_all}
\end{figure}

To make an accurate comparison with the observed cluster mass function, we need to consider the relative number of GMCs with different mass. The powerlaw index for the GMC mass distribution in the 
inner Milky Way is approximately -1.5 \citep{Sanders,Solomon1987,Roso2005}. Taking the GMC mass distribution to be

\begin{equation}
\frac{dN}{dM} \propto M^{-1.5}
\end{equation}

\noindent in the range 10$^4$ - 10$^6$ M$_{\odot}$, we combine the cluster data from our 5 RHD GMCs at 5 Myr, weighted by the relative numbers of the 
clouds in which they were born

In Figure \ref{fig:ECMF_all} we show the resulting, computed cluster mass function that arises from the Milky Way cloud mass function. This can then be compared to the observed mass 
distribution which is a collection of distinct clusters in forming in different regions. Note that we only include the stellar mass of each cluster when
producing this distribution. The Figure shows that the cluster mass function peaks at $\sim$10 M$_{\odot}$ corresponding to a small stellar group. The peak 
cluster mass is on the same order of magnitude as the 50 M$_{\odot}$ turnover found in \citet{Lada2003}. 

\begin{figure*}
\includegraphics[width=0.98\linewidth]{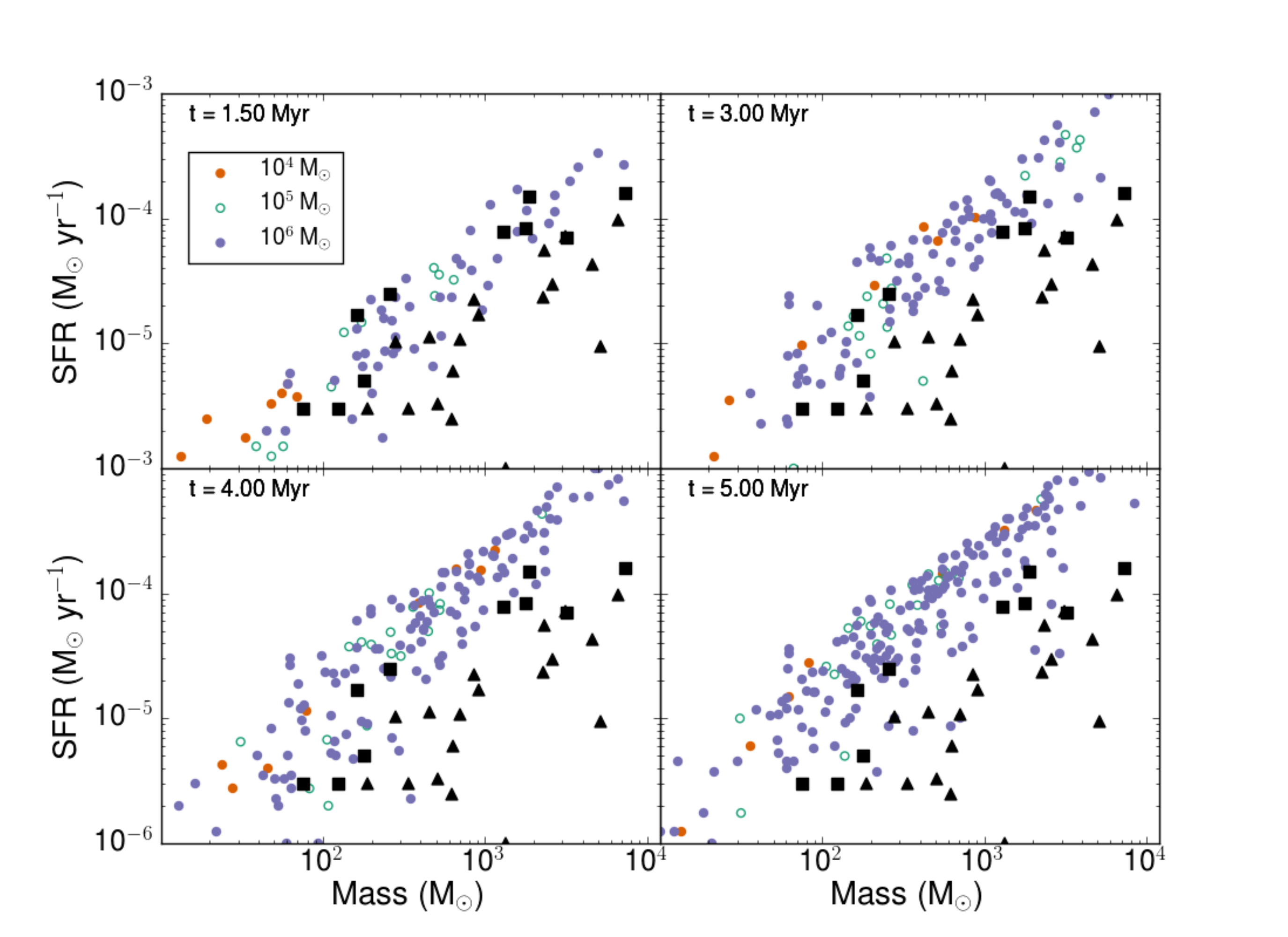} 
\caption{The SFR of individual cluster particles at various times for the 10$^4$, 10$^5$, and 10$^6$ M$_{\odot}$
GMCs. The squares are the observational results from \citet{Lada2010} and the black triangles are a similar data set from \citet{Heidermann2010}.}
\label{fig:SFR_all}
\end{figure*}

We make a further comparison to the results of \citet{Lada2003} who measured the embedded cluster mass function of nearby star-forming regions and found $dlog(N)/dlog(M)\propto M^{\beta}$, where $\beta$ $\sim$ -1. 
The cluster mass function has been measured for extragalactic clusters (see \citet{FallChandar} for comprehensive overview) and the same functional form is also found which, in some galaxies, 
extends to $>$10$^5$ M$_{\odot}$ clusters. Embedded clusters of this mass are not seen in the MW.

Motivated by the observational data and a cluster mass function which appears consistent with a broken powerlaw, we provide two fits to our data --- one covering the range of 
observed, embedded clusters in the MW (solid line), and one for the higher mass clusters (dashed line). We only include cluster masses greater than 10 M$_{\odot}$ in the calculation.

Fitting over the range of embedded clusters in the MW --- 1 $\leq$ log(M/M$_{\odot}$) $\leq$ 3.3 \citep{Lada2003} --- results in a slope of $\beta$ = -0.99 $\pm$ 0.14. 
This is consistent with the observed slope. Performing the same analysis for the HD simulations, we find a slope of -0.83 that is significantly shallower than observed. This clearly shows that 
radiative feedback is playing some role in limiting the growth of clusters since there are a relatively larger amount of high mass clusters in the HD simulations.

A slope of $\beta$ = -2.82 is found for cluster masses greater than $\sim$2000 M$_{\odot}$. One reason for the steeper slope at high masses is that the largest GMCs --- the source of the most massive clusters --- are not disrupted at the end of 
the 5 Myr simulation. This suggests that these clusters will continue to grow and accrete gas, leading to more clusters populating the high mass end of the distribution.  

\subsection{Star Formation Rates}

We plot the SFRs of our individual clusters versus the cluster mass at various times in Figure \ref{fig:SFR_all}. We only show the 10$^4$, 10$^5$ and 10$^6$ M$_{\odot}$
for clarity.

We have over plotted the results from \citet{Lada2010}, shown by black squares, who measured the SFRs of local star forming regions by counting Young Stellar Objects (YSOs)
and adopting a star formation timescale to estimate the SFR. In order to make an accurate comparison to these results, we adopt the same model parameters as \citet{Lada2010}
and estimate the SFR via recently formed stars. We direct the reader to \citet{Howard1} which plotted the SFRs in the same way and contains more detail about this procedure. We also include the results from \citet{Heidermann2010} who 
also measured the SFRs of local regions, some of which are also included in the \citet{Lada2010} dataset.

Our simulated SFRs agree well with the observed values at early times, particularly with the \citet{Lada2010} results. At 3 Myr, the low to intermediate mass clusters are still
consistent with the measured SFRs, but the high mass clusters are overproducing stars. This is strong evidence that radiative feedback alone is not sufficient for limiting
the SFR (and therefore the SFE) since these high mass clusters have the highest ionizing luminosity and should be influencing their surroundings the strongest.

The trend of high SFRs extends to all mass regimes past 4 Myr. While the slope of the SFR-mass relation is consistent with the observations, the normalization is not. This
is also true for the 10$^5$ M$_{\odot}$ GMC which, as shown in Section 3.2, had a large reduction in $\epsilon_{cl}$ when including radiative feedback and a globally decreasing
SFR at late times. This also supports the claim that other feedback mechanisms, such as stellar winds, are required to explain the SFRs of young, nearby star-forming regions \citep{Gatto2017}.

We note that high SFRs may also be due, in part, to our adopted subgrid model for star formation. We do not include feedback on scales smaller than our cluster particles, and so
any accreted gas will inevitably be converted to stars over a long enough timescale. We refer the reader to Section 2.1 of \citet{Howard1} for a detailed discussion of this point.

If our cluster SFRs are artificially high, the total luminosity will also be too high. This means that the impact radiative feedback has on each cloud, as discussed above,
should be considered an upper limit. This lends further support for the need of other feedback mechanisms during the early phases of cluster formation since the maximum
suppression of $\epsilon_{cl}$ relative to HD simulations was approximately a factor of two.

\section{Discussion and Conclusions}

We examine the early phases of cluster formation and the role of radiative feedback in a suite of GMCs that have masses in the range of 10$^{4-6}$ M$_{\odot}$. To isolate
the role of GMC mass, we use the same initial density and virial parameter across clouds. Sink particles are used to represent the formation of a cluster and a custom 
subgrid model is used for star formation within the clusters. The properties of the stellar population in each cluster is tracked and, combined with a raytracing
radiative transfer scheme, is used to compute the radiative feedback. 

The main result of this work is that the strength of radiative feedback depends on the initial GMC mass. The fractional reduction in the cluster formation efficiency, $\epsilon_{cl}$, when including
radiative feedback is the largest for the 5$\times$10$^4$ and 10$^5$ M$_{\odot}$ GMCs. Both of these models had $\epsilon_{cl}$ reduced by a factor of $\sim$2 relative to purely hydrodynamical simulations.
The star formation efficiency in these clouds was reduced by 30-40\%. In contrast, the lowest mass model (10$^4$ M$_{\odot}$) showed a only $\sim$20\% reduction in $\epsilon_{cl}$. 

The variation in the impact radiative feedback has on the cluster and star formation efficiencies is attributed to the balance between how much radiation energy is absorbed by the
GMC and the gravitational potential energy of the cloud. The smallest GMC is not massive enough to form a population of massive stars and therefore cannot effectively limit
early star and cluster formation. The highest mass objects, on the other hand, do produce massive stars but their corresponding gravitational potential is too large for the cloud
to be globally unbound. The regime between these two limits (5$\times$10$^4$ to 10$^5$ M$_{\odot}$) balances these two effects, leading to a larger suppression in $\epsilon_{cl}$ and 
$\epsilon_{*}$. We have shown this by plotting the ratio of the internal energy, a proxy for the amount of absorbed radiation energy injected by the star-forming 
clusters, to the gravitational potential energy of the cloud. This ratio exceeds one, indicating the cloud has been globally unbound, for the 5$\times$10$^4$ and 
10$^5$ M$_{\odot}$ clouds at 3.5 and 3.8 Myr. respectively. The higher mass GMCs always have a ratio below 1, indicating that gravity dominates, and the ratio for the 10$^4$ M$_{\odot}$
only exceeds 1 at late times at which point the majority of star and cluster formation has already occurred.

The other important conclusions of this work can be summarized as follows:

\begin{itemize}

\item The cluster formation efficiency ($\epsilon_{cl}$) and the star 
formation efficiency ($\epsilon_{*}$) vary significantly across different mass GMCs. At 5 Myr, $\epsilon_{cl}$ is 43\%, 29\%, 23\%, 35\%, and 28\% for the 10$^4$, 5$\times$10$^4$, 
10$^5$, 5$\times$10$^5$, and 10$^6$ M$_{\odot}$ GMCs, respectively. At the same time, $\epsilon_{*}$ ranges from 16\%-21\%.
 
\item The high SFEs found in all models, even when including radiative feedback, suggests that other forms of feedback, such as stellar winds, are required to limit
early star formation in GMCs \citep{Gatto2017}. This is further supported by the comparison between our clusters and the SFRs of local star-forming regions. We find good agreement with observed
SFR-mass relation at early times, but by $\sim$3 Myr our clusters are systematically overproducing stars. The slope of our SFR-mass relation, however, is consistent with observed star-forming clusters at all times.

\item We produced an initial cluster mass function by combining the results from all RHD simulations weighted by the Galactic GMC mass function. The resulting slope of 
the powerlaw distribution ($dlog(N)/dlog(M) \propto M^{\beta}$) over the range of embedded cluster masses in the MW (log(M/M$_{\odot}$) $<$ 3.3) is $\beta$ = -0.99$\pm$0.14. This is consistent
with the observed slope. Performing the same analysis for the HD simulations results in a shallower slope of -0.83, indicating that radiative feedback 
does contribute to limiting the growth of star-forming clusters. An apparent break in the powerlaw above log(M/M$_{\odot}$) = 3.3 is attributed to the 5 Myr timescale of our simulations. The most massive GMCs are not yet disrupted and, given more time, will fill out the high mass end of the distribution.

\item The limiting of cluster growth by radiative feedback is also supported by the relation between the host GMC mass (M$_{GMC}$) and the maximum mass cluster it forms (M$_{c,max}$). Using the RHD data, we find that
M$_{c,max}\propto$ M$_{GMC}^{0.81}$. The HD simulations always form higher mass clusters and the relation is instead M$_{c,max}\propto$ M$_{GMC}^{0.93}$. The steeper slope for the
HD simulations indicates that the largest clusters in the highest mass GMCs are more strongly limiting their growth via radiative feedback compared to those formed in smaller clouds.
\end{itemize}

\section*{Acknowledgments}

We thank the anonymous referee for useful suggestions. C.S.H. acknowledges financial support provided by the Natural Sciences and Engineering
Research Council (NSERC) through a Postgraduate
scholarship. R.E.P. and W.E.H. are supported by Discovery Grants from the
Natural Sciences and Engineering Research Council
(NSERC) of Canada. The FLASH code was in part developed
by the DOE supported Alliances Center for Astrophysical
Thermonuclear Flashes (ASCI) at the University of
Chicago. Computations were performed on the gpc supercomputer at the SciNet HPC Consortium. SciNet is funded by: the Canada Foundation 
for Innovation under the auspices of Compute Canada; the Government of Ontario; Ontario Research Fund - Research Excellence; and the University of Toronto.

\bibliography{howardpaper}
\bsp

\label{lastpage}

\end{document}